\documentclass{emulateapj}
\usepackage{apjfonts}
\usepackage{times} 
\usepackage{graphicx}
\usepackage{rotating}
\usepackage{array}
\usepackage{epsfig}
\usepackage{amsmath}
\usepackage{natbib}
\def\mbh{\ifmmode{{\mathrm M}_{bh}\,}\else{M$_{bh}$\,}\fi}
\def\msigma{\ifmmode{{\mathrm M}_{bh}-{\sigma}\,}\else{M$_{bh}- \sigma$\,}\fi}
\def\msun{\ifmmode{{\mathrm M}_{\odot}}\else{M$_{\odot}$\,}\fi} 
\def\kms{\ifmmode{{\mathrm{km \, s^{-1}}}}\else{${\mathrm{km \, s^{-1}}}$}\fi} 
\def\kmskpc{km~s$^{-1}$~kpc$^{-1}$}

\shorttitle{ Triaxial Galaxies with Figure Rotation}
\shortauthors{Deibel, Valluri, \& Merritt }

\begin{document}
\title{The Orbital Structure of Triaxial Galaxies with Figure Rotation}

\bigskip

\author{Alex T. Deibel,\altaffilmark{1,2}
Monica Valluri,\altaffilmark{1}
and David Merritt,\altaffilmark{3}
}
\altaffiltext{1}{Department of Astronomy, University of Michigan, Ann Arbor, MI 48109, USA {\tt mvalluri@umich.edu}}
\altaffiltext{2}{Department of Physics and Astronomy, Michigan State University, East Lansing MI, 48824, USA}
%, {\tt deibelal@msu.edu}}
\altaffiltext{3}{Department of Physics and Center for Computational Relativity and Gravitation, Rochester Institute of Technology, 84 Lomb Memorial Dr., Rochester, NY 14623, USA}

%\date{ \today}

\begin{abstract} 
We survey the properties of all orbit families in the rotating frame
of a family of realistic triaxial potentials with central supermassive
black holes (SMBHs). In such galaxies, most regular box orbits (vital for
maintaining triaxiality) are associated with resonances which occupy
two-dimensional surfaces in configuration space.  For slow figure
rotation all orbit families are largely stable.  At intermediate
pattern speeds a significant fraction of the resonant box orbits as
well as inner long-axis tubes are destabilized by the ``envelope
doubling'' that arises from the Coriolis forces and are driven into
the destabilizing center.  Thus, for pattern rotation periods $2\times
10^8{\rm yr} \lesssim T_p \lesssim 5\times 10^9{\rm yr}$, the two orbit families
that are most important for maintaining triaxiality are highly
chaotic. As pattern speed increases there is also a sharp decrease in
the overall fraction of prograde short-axis tubes and a corresponding
increase in the retrograde variety.  At the highest pattern speeds
(close to that of triaxial bars), box-like orbits undergo a sudden
transition to a new family of stable retrograde loop-like orbits, which
resemble orbits in three-dimensional bars, and circulate about the short axis.  Our
analysis implies that triaxial systems
(with central cusps and SMBHs) can either have
 high pattern speeds like fast bars or low patten speeds like triaxial elliptical galaxies or dark matter halos found in $N$-body simulations.
Intermediate pattern speeds produce a high level of stochasticity in both the box and inner long-axis tube orbit families implying that stable triaxial systems are unlikely to have such pattern speeds.

\end{abstract}

\vspace{1.cm}

\section{Introduction}
\label{sec:intro}

It is widely accepted that since both elliptical galaxies and dark
matter halos form via hierarchical mergers, they should be
triaxial. If a significant amount of gas is present in the progenitor
galaxies, its dissipative condensation generally results in systems
that are more oblate than those produced in purely collisionless
collapse \citep{dubins_94, kkzanm04, deb_etal_08}. However, recent
studies have shown that two orbit families that characterize triaxial
systems (boxes and long-axis tubes) persist in significant numbers
even when a triaxial system {\em looks} almost oblate
\citep{vandenbosch_dezeeuw_10,valluri_etal_10}. Since merging systems generally have angular momentum (either internal angular
momentum of the individual progenitors or the angular
momentum of their relative orbit or both), merger remnants generally have angular momentum.  If the remnants are triaxial, they can
exhibit figure rotation: a property that is independent of the
streaming motions of individual particles.

Although it is logical to assume that triaxial elliptical galaxies can
have figure rotation, it is currently difficult (if not impossible) to
observationally distinguish between figure rotation (tumbling) and
orbital streaming. Although figure rotation was first proposed to
explain ``anomalous dust lanes'' in triaxial elliptical galaxies
\citep{vanalbada_etal_82}, there have been only a few observational
attempts to measure the pattern speeds of either elliptical galaxies
or dark matter halos \citep{bureau_etal_99,
  jeong_etal_07}. Cosmological $N$-body simulations (without
dissipation) predict that dark matter halos are significantly triaxial
and that a majority of dark matter halos ($\sim$90\%) have figure
rotation \citep{bailin_steinmetz_04,bryan_cress_07} with slow pattern
speeds following a lognormal distribution centered on $\Omega_p =
0.148h$\kmskpc.  \citet{bailin_steinmetz_04} found that the pattern
speed of the figure rotation is correlated with the cosmological halo
spin parameter $\lambda$ \citep{peebles_69}, but is independent of
halo mass. Figure rotation of triaxial dark matter halos has been
suggested as a mechanism for driving spiral structure, warps and bars
in spiral and elliptical galaxies with low mass disks
\citep{bureau_etal_99, dubinski_chakrabarty_09}.  The only family of
triaxial stellar systems with strong figure rotation is the bars of
spiral and lenticular galaxies. Rapidly rotating (triaxial) bars can
be almost completely regular \citep{pfenniger_friedli_91} and have a
complex orbital structure
\citep{skokos_etal_02a,skokos_etal_02b,patsis_etal_02,harsoula_kalapotharakos_09}.

\citet{emsellem_etal_07} introduced a luminosity weighted measure of
the specific line-of-sight angular momentum, $\lambda_R$, to quantify
the angular momentum content of E/S0 galaxies observed in the SAURON
sample. They found that this parameter allows E/S0 galaxies to be
classified into two subgroups $-$ the ``fast rotators'' with $\lambda_R
> 0.1$ and ``slow rotators'' with $\lambda_R < 0.1$. The slow rotators
are likely to be triaxial ellipticals since they frequently display
isophotal and kinematic twists, including kinematically decoupled
cores. They tend to be more luminous and have shallow central cusps
\citep{emsellem_etal_07, cappellari_etal_07}. In contrast, ``fast
rotators'' are more numerous and close to axisymmetric, but may retain
a significant population of triaxial orbit families such as box and
long-axis tube orbits \citep{vandenbosch_dezeeuw_10}. These authors
find that even a small fraction of box-like orbits can significantly
alter the dynamical estimates of supermassive black hole (SMBH)
masses. Recent $N$-body simulations of mergers of gas rich disk
galaxies that include star formation and dissipation show that both
fast and slow rotators can be formed in mergers between disk galaxies.
The latter class primarily form in major mergers with smaller gas
fractions and are more triaxial \citep{jesseit_etal_09}. (Recently,
\citet{bois_etal_10} have shown that due to numerical resolution
effects in $N$-body simulations, the fraction of merger remnants that
are triaxial slow rotators may have been underestimated). Thus despite
the absence of concrete observational measurements, some fraction of
triaxial elliptical galaxies and dark matter halos could have figure
rotation.

There has been little work on the effects of figure rotation on
triaxial galaxies in the last two decades $-$ none on realistic density
profiles with central cusps and SMBHs. \citet{binney_81} studied the effect of figure rotation on the
behavior of the closed periodic orbits in the equatorial plane
perpendicular to the axis of rotation ($z$) of flattened galaxy models
including those with weak bars.  He showed that %the equations of
%motion could be reduced to the Mathieu equation and found that 
closed retrograde orbits in the $x$-$y$ plane lying in an annular region
(now called the ``Binney instability strip'') become unstable to
perturbations perpendicular to the plane. He also showed
that this instability strip moved inward as the speed of rotation of
the figure increased.

%Since the family of closed periodic orbits that lie in the $x-y$
%plane ``parents'' the short axis tube family this suggests that
%retrograde short axis tubes are likely to become unstable as figure
%rotation increases. We will show in \S~\ref{sec:results} that this is
%not found to be true for the full orbit population explored in our
%models with cusps and central black holes.

\citet{heisler_etal_82} studied the stability of closed orbits in a
triaxial model with a central density core subjected to both slow
(pattern rotation period $T_p \sim 10^9$~yr) and fast rotations ($T_p
\sim 10^8$~yr).  They showed that the 1:1 periodic orbits that
circulate about the long ($x$) axis of the model are stable to figure
rotation and are tipped about the $y$-axis by the Coriolis forces in a
direction that depends on their sense of motion. Two such orbits
exist: one rotates clockwise about the $x$-axis and the other
circulates anti-clockwise. However, both ``tipped'' orbits circulated
about the short axis in the same direction. (These orbits were termed
``anomalous'' by \citet{vanalbada_etal_82}.)  
%\citet{heisler_etal_82} also showed that this series of closed 
%orbits terminates at the outer edgeof Binney instability strip.  
The long-axis tube family that is ``parented'' by the anomalous orbits
was therefore also expected to be stable and also ``tipped'' at an
angle. They noted that for orbits with very large energies
(i.e., extending to large radii) such orbits could be tipped by $\sim
90^\circ$ into the equatorial plane. Our more general analysis in
models with realistic density profiles shows that while outer
long-axis tubes indeed behave as predicted, the inner long-axis tubes
with the small pericentric radii are easily destabilized by figure
rotation (Section~\ref{sec:tubes}).

%\citet{binney_81,heisler_etal_82} also found that the 1:1 closed
%periodic orbit that circulates the intermediate-axis is unstable both
%in non-rotating and rotating models, accounting for the absence of
%intermediate axis tubes in self-consistent numerical models of
%triaxial galaxies examined by \citet{heiligman_schwarzschild_79}.
\citet{heisler_etal_82} also showed that the normal retrograde orbits
that lie in the equatorial plane were stable (except in the Binney
instability strip, where they were unstable to vertical
perturbations).  The sequence of closed, stable period orbits
identified by them were found to exist at both slow and fast pattern
speeds and constituted one composite sequence which was stable over
the entire energy range. They speculated that this implied that
triaxial galaxies could have both fast and slow pattern speeds.

\citet{dezeeuw_merritt_83} complemented this work, with a study of
orbits in the principal planes of a rotating triaxial galaxy with a
central core. They found three prograde sequences of stable orbits in
addition to the retrograde sequences.
%Outside the core they showed that 1:1 periodic orbits in $x-y$ 
%and $y-z$ planes were stable and would also be tipped about the $y$-axis. 
Inside the core of the galaxy, they found that the $x$-axial orbit was
stable and generated a family of box orbits that underwent ``envelope
doubling'' as they looped around the center due to the Coriolis
forces.
 
The first attempt to construct a self-consistent triaxial galaxy with
slow figure rotation was presented in the pioneering work of
\citet{schwarzschild_82}. Based on previous studies of periodic
orbits, Schwarzschild restricted his rotating triaxial models to a
moderate pattern speed ($T_p\sim 1.2\times 10^9$yr) for which the main
resonances (corotation, outer inner Lindblad, and the Binney
instability strip) lay outside the model. He showed that self-consistent
triaxial models with figure rotation could be constructed such that
they had either no streaming motions or maximal streaming.
%He found that the streaming motions had a complicated pattern, 
% prograde in the inner regions and retrograde at larger radii. 

The main conclusion from the early work on figure rotation in triaxial
galaxies with cores was that except for the Binney instability
associated with retrograde orbits confined to the equatorial plane,
several stable sequences of orbits that parented the major orbit
families existed over a range of pattern speeds and energies, implying
that triaxial galaxies with both fast (rotation period $T_p \sim
10^6$~yr) and moderate rotation speeds ($T_p \sim 10^9$~yr) could exist. Our
findings here will show that the presence of a central cusp or SMBH alters that conclusion.

\citet{gerhard_binney_85} first proposed that the box orbits which
form the back bone of triaxial elliptical galaxies would become
chaotic due to scattering by the divergent central force arising from
a central black hole. The presence of a significant fraction of chaotic orbits results in chaotic mixing which can cause secular relaxation of orbits in phase space \citep{kandrup_mahon94} resulting in  a change in the shape of the galaxy from triaxial to axisymmetric on timescales of order $\sim 30$-$50$ dynamical times \citep{merritt_valluri_96,merritt_quinlan_98}.  \citet{gerhard_binney_85} had also argued, however, that if the triaxial
potential had figure rotation, then box orbits  (which are crucial to maintaining triaxiality) would be less affected
by the central force in the rotating frame due to ``envelope
doubling'' \citep{dezeeuw_merritt_83}.
%\citet{gerhard_binney_85} suggested that this would cause the
%$x$-axial orbits (and therefore all box orbits ``parented'' by the
%$x$-axial orbit) to be less affected by chaotic scattering.

Several early studies indicated that triaxial galaxies with figure
rotation could often have significant fractions of chaotic orbits.  The
studies by
\citep{udry_pfenniger_88,martinet_udry_90,udry_91,tsuchiya_etal_93}
were however restricted to limited numbers of orbits in models with
cores and did not pursue the primary cause of this chaos.

It is now believed that all elliptical galaxies have either shallow or steep central density
cusps \citep{gebhardt_etal_96,lau_etal_07} and central SMBHs \citep{ferrarese_merritt_00,gebhardt_etal_00}. In
realistic elliptical galaxy models with cusps and black holes, a large
fraction of phase space is occupied by resonant and chaotic orbits
\citep{miralda_escude_schwarzschild_89,
  schwarzschild_93,valluri_merritt_98, merritt_valluri_99,
  poon_merritt_04}. Although the fraction of chaotic orbits increases
with an increase in the strength of the density cusp
\citep{schwarzschild_93, merritt_fridman_96,merritt_97} or the mass of
the central black hole \citep{merritt_valluri_96}, it is still
possible to construct triaxial models that do not evolve rapidly due
to chaotic mixing so long as a significant fraction of orbits ($\sim
50$\%) are regular \citep{poon_merritt_02, poon_merritt_04}.

In a precursor to the present paper, \citet{valluri_99} first showed
that box orbits in triaxial galaxies with cusps and black holes are
destabilized by moderate amounts of figure rotation because the
envelope doubling acts to further destabilize the resonant box orbits,
rather than stabilize chaotic orbits as predicted by
\citet{gerhard_binney_85}. This paper further explores the cause of
the destabilization of box-like orbits and investigates the effects of
figure rotation on all major orbit families.

Our objective in this paper is to study the behavior of all the major
and minor orbit families, especially those that form the backbone of
realistic rotating triaxial galaxy models. The goal is to identify the
range of rotation speeds for which such orbits will remain stable.
%We find that in realistic galactic potentials there are two restricted ranges of pattern speeds for which the majority of orbits will remain stable.

The paper is organized as follows. In Section~\ref{sec:method}, we describe
the numerical model, the selection of orbital initial conditions, and
the Laskar mapping method \citep{laskar_90}.  In Section~\ref{sec:results},
we describe the results of our analysis of the effects of figure
rotation on major orbit families, as a function of pattern speed,
orbital energy, and the shape of the triaxial model, in models with
and without central SMBHs. We summarize our results
and discuss the implications of our findings in Section~\ref{sec:discuss}.

\section{Models and Numerical Methods}
\label{sec:method}

\subsection{Density model}
\label{sec:model}

We studied triaxial generalizations of spherical models first
presented by \citet{dehnen_93} and \citet{tremaine_etal_94}. The model, which we
will henceforth refer to as the ``Dehnen-model'', has a density law
that is a good fit to the observed luminosity profiles of ellipticals
and the bulges of spirals, and is given by
\begin{equation}
\rho(m) = {(3-\gamma) M\over 4\pi abc} m^{-\gamma} 
(1+m)^{-(4-\gamma)},
\ \ \ \ 0 \le \gamma < 3,
\label{deh1}
\end{equation}
where
\begin{equation}
m^2={x^2\over a^2} + {y^2\over b^2} + {z^2\over c^2}, \ \ \ \ 
a\ge b\ge c\ge
0,
\label{deh2}
\end{equation}
and $M$ is the total mass of the model. The mass distribution is
stratified on concentric ellipsoids where $a, b,$ and $c$ are the scale
lengths of the semimajor, semi-intermediate, and semiminor axes of
the model, which are aligned with Cartesian coordinates $x, y,$ and $z$,
respectively.  The parameter $\gamma$ which determines the logarithmic
slope of the central density cusp ranges observationally from $\gamma
= 0.5$-$1$ in luminous galaxies with ``shallow cusps'' (sometimes
called a ``core'') to $\gamma=2$ in the lower luminosity galaxies with
``steep cusps'' \citep{gebhardt_etal_96,lau_etal_07}. At large radii the density
profile of the model always falls as $m^{-4}$. The model has a finite
density core for $\gamma = 0$ and an infinite central density for
$\gamma > 0$.  The potential ($\Phi(\mathbf{x})$) and forces in the
stationary triaxial Dehnen model in ellipsoidal coordinates are taken
from \citet{merritt_fridman_96}.

In the rotating frame, the energy of an orbit is not an integral of motion but the Jacobi integral ($E_J$) is a conserved quantity:
\begin{equation}
E_J = {1\over 2}|\dot{\mathbf{x}}|^2+\Phi-{1\over 2}|\Omega_p \times {\mathbf{x}}|^2,
\end{equation}
where $\mathbf{x}$ and $\mathbf{\dot{x}}$ are three-dimensional spatial and velocity vectors, respectively. 

In our models, figure rotation is about the short ($z$) axis, hence $\vec{\Omega}_p= \Omega_p \hat{e_z}$ and equations of motion in the rotating frame (BT08, Section~3.3.2) become
\begin{align}
\mathbf{\ddot{x}} &=    -\nabla{\Phi} - 2(\vec{\Omega}_p\times{\mathbf{\dot{x}}}) -\vec{\Omega}_p\times(\vec{\Omega}_p\times{\mathbf{x}})\\
                            &=   -\nabla{\Phi} - 2(\vec{\Omega}_p\times{\mathbf{\dot{x}}}) +|\Omega_p|^2{\mathbf{x}}.
\end{align}
In Cartesian coordinates the equations of motion are then given by
\begin{align}
\ddot{x}  =&  - {{\partial\Phi\over{\partial x}}} - 2\Omega_p y+ \Omega_p^2 x,\\
\ddot{y}  = & - {{\partial\Phi\over{\partial y}}} +2\Omega_p x+ \Omega_p^2 y,\\
\ddot{z}  =  &- {{\partial\Phi\over{\partial z}}}, 
\end{align}
where $-2\Omega_p y$ and $2\Omega_p x$ are Coriolis force terms and $\Omega_p^2x$ and $\Omega_p^2y$ are centrifugal force terms.\footnote{Following current convention (e.g., BT08) we use the right-handed screw rule (with positive angular momentum vector pointed up) for figure rotation and a right-handed coordinate system. Note that some previous authors \citep{schwarzschild_82} used a left-handed screw rule for figure rotation and a right-handed coordinate system.}

%The density is derived from the potential 
%\begin{equation}
%\Phi(r) =  
%\begin{cases} -\frac{GM}{(2-{\gamma})a}\left[1-\frac{(r/a)^{2-{\gamma}}}{(1+r/a)^{2-{\gamma}}}\right],   & \gamma \ne 2
%\\
%\\
%-\frac{GM}{a}log(1+\frac{a}{r}),  & \gamma = 2 
% \end{cases} 
%\label{deh3}
%\end{equation}
%via Poisson's equation. This potential yields a radial force of
%\begin{equation}
%-\frac{\partial{\Phi}}{\partial{r}} = -\frac{GM}{a^2}\left(\frac{r}{a}\right)^{1-\gamma}\left(1+\frac{r}{a}\right)^{\gamma - 3},
%\label{deh4}
%\end{equation}
%as shown in Dehen (1993).  

%The radial force is finite at the center for $\gamma < 1$ (``weak
%cusp'') and divergent for $\gamma \ge 1$ (``strong cusp''). We will
%take $\gamma = 1$ to be a realistic value for cusp slope for
%elliptical galaxies with ``shallow cusps and $\gamma = 2$ corresponds
%to galaxies with steep cusps. The Dehnen model is a reasonable
%description for stellar densities at small and intermediate radii in
%elliptical galaxies.  

Most of the models used in this study are close to maximally triaxial
with triaxiality parameter $T= (a^2 - b^2)/(a^2 - c^2) =0.58$ and
minor to major axis ratio $ c/a=0.5$. For a limited number of models,
we also explored the effect of changing the shape of the triaxial
figure with $c/a = 0.5, 0.8,$ and $0.7$ with triaxiality parameters $T= 0.1,
0.9,$ and $0.3$, respectively.  The shapes therefore range from nearly
oblate, through oblate triaxial to nearly prolate.

Following the standard practice, we adopt a set of units where the
total galactic mass $M$, the semimajor axis scale length $a$, and the
gravitational constant $G$ are set to unity.  When a central black
hole is added to the model, its mass $M_{bh}$ is expressed as a
fraction of the total galaxy mass $M$. In this paper we restrict
ourselves to studying models with either no central point mass ($M_{bh}
= 0$) or models with $M_{bh}=0.001$ (i.e., 0.1\% of the mass of the
galaxy). The latter value is consistent with $0.14\%\pm 0.04\%$ the
mass fraction in a central SMBH that is expected to
be found in most elliptical galaxies \citep{haring_rix_04}.  The
potential and forces due to the central black hole are those of a
softened point mass with softening length $\epsilon = 10^{-5}a$.

The orbital structure of the stationary versions of triaxial Dehnen
models with various cusp slopes and a range of black hole mass
fractions have been previously studied
\citep{merritt_fridman_96,valluri_merritt_98,siopis_kandrup_00}. In
this paper we restrict ourselves to presenting models with $\gamma =
1$ for the following reasons.  First, the most likely candidates for
triaxial elliptical galaxies with slow figure rotation are the more
luminous ``slow rotators'' with shallow central cusps
\citep{emsellem_etal_07}.  Second, a Dehnen model with $\gamma=1$ is
quite similar (at least in the inner regions) to the density profiles
of cosmological dark matter halos with the main difference being that
cosmological density profiles fall off more slowly at large radii
($r^{-3}$ compared to $r^{-4}$ for the Dehnen model). 
%Furthermore, the
%differences in the slopes at large radii may in fact be transient,
%since it has been shown that cosmological simulations evolved into the
%future result in steeper asymptotic outer density profiles of $r^{-4}$
%\citep{busha_etal_07}.  
Models with $\gamma=2$ were studied (but are
not presented) since their dependence on pattern speed is
 qualitatively similar to that of the $\gamma=1$ models with an SMBH. 
 (We note that while our choice of triaxial potential is
representative of a triaxial elliptical galaxy, it is an incomplete representation of a dark matter halo which could have a significant potential contribution from a stellar disk.)

At the present time only a few measurements of the pattern speeds of
figure rotation in triaxial dark matter halos and early-type galaxies
exist.  The pattern speeds of fast bars measured by applying the
Tremaine$-$Weinberg method \citep{tre_wei_84,tremaine_weinberg_meth_08}
typically indicate that the ratio of the corotation radius to the length
of the bar $R_\Omega/a = [1, 1.4]$ (for a bar of semi-major axis
length $a$) \citep[e.g.,][]{debattista_etal_02,aguerri_etal_03,
  corsini_10,BT08}. Hereafter we shall use the quantity $R_\Omega/a_i$
to compare the pattern speeds of orbits launched from different radial
shells (of semimajor axis length $a_i$) to each other and to the
pattern speeds of bars. 

The pattern speed of early-type galaxy NGC~2974 has been measured by
fitting the properties of three rings of recent star formation
\citep{jeong_etal_07}. This galaxy is normally classified as E4, but
the authors argue that all three rings can be accounted for as
occurring at resonances if this galaxy is a S0 galaxy with an extended
stellar bar and pattern speed, $\Omega_p \sim 78\pm6$~\kmskpc. This
pattern speed is only slightly slower than that of fast bars. Figure
rotation of a triaxial dark matter halo was proposed as the cause of
the extended spiral arms in the blue compact dwarf galaxy NGC 2915
\citep{bureau_etal_99, bekki_freeman_02} whose modeling required a
halo pattern speed of $\Omega_p= 7\pm1$~\kmskpc. This pattern speed is
an order of magnitude larger than the maximum pattern speed measured for dark matter halos produced in cosmological $N$-body simulations
($\Omega_p=1.01h$~\kmskpc)  and nearly two orders of magnitude larger than the median pattern speed ($\Omega_p=0.148h$~\kmskpc) \citep{bailin_steinmetz_04}. Thus, at the present time, observational constraints on the pattern speeds of triaxial halos and elliptical galaxies are quite uncertain.  Our fastest pattern speed $\Omega_p=34$~\kmskpc (for $R_\Omega/a = 2$) is slightly slower than the speeds observed in fast bars, our slowest $\Omega_p=0.58$~\kmskpc (for $R_\Omega/a = 40$) is comparable to fastest measured in simulations of dark matter halos. We explore a range of values for $\Omega_p$ between these limits. 

The triaxial Dehnen models studied here rotate about the short axis
($z$). Simulations of collisionless dark matter halos show that the
angular momentum axes have a mean misalignment of $\sim 25^\circ$ with
the minor axis \citep{bailin_steinmetz_05}. However, the dissipative
collapse accompanying galaxy formation is likely to induce angular
momentum transport and a higher degree of alignment between the spin axis and the short axis of the galaxy. The pattern speed
of the triaxial figure is given in terms of the ``corotation radius'',
hereafter $R_\Omega$. In a nearly axisymmetric potential, the corotation radius is the radius at which
the   frequency $\Omega_c$ of a closed (almost circular) orbit in the equatorial
($x$-$y$) plane of the potential is the same as the pattern frequency
(generally called ``pattern speed'') $\Omega_p$:
\begin{equation}
%\Omega_p =\Omega_c =  {\sqrt{{\frac{1}{R}}\big(\frac{\partial \Phi}{\partial R}\big)}}\vert_{R=R_\Omega}.
\Omega_p =\Omega_c =  \Big[{\sqrt{{\frac{1}{R}}\nabla{\Phi}}}\Big]_{R=R_\Omega, \, z=0}.
\end{equation}

Working in the equatorial plane of our triaxial model we set $R^2=(x/a)^2+(y/b)^2$. 
In our models $R_\Omega$ is given in units of $a$, the scale length of
the semimajor axis, and ranges from $R_\Omega = 60$ (slowly rotating)
to $R_\Omega = 2$ (very rapidly rotating) (see
Table~\ref{tab:rotation}). In this nomenclature the stationary
(non-rotating) model has its corotation radius at infinity and is labeled
with $R_\Omega = \infty$. We do not discuss the orbital structure of
more prolate triaxial structures similar to fast bars with higher
pattern frequencies ($R_\Omega \leq 2$) since they have been
previously studied \citep{skokos_etal_02a,skokos_etal_02b,
  harsoula_kalapotharakos_09}.  To give the reader a better physical
appreciation of the pattern frequencies implied by the co-rotation
radii, we convert our model units to physical units in an elliptical
galaxy. For a semi-major axis scale radius $a=5$~kpc and a galaxy mass
of $M=5\times10^{11}M_{\odot}$, the unit of time for the model is given
by
\begin{equation}
T = \sqrt{\frac{a^3}{GM}} = 1.49\times 10^6 \left[\frac{a}{\rm kpc}\right]^{3/2} \left[\frac{M}{10^{11}M_{\odot}}\right]^{-1/2} {\rm yr}. 
\end{equation}
For these parameters Table~\ref{tab:rotation} gives the pattern
frequency ($\Omega_p$) and rotation period $T_p = 2\pi/\Omega_p$ (in
years) for each value of corotation radius $R_\Omega$ that was
studied.

\begin{table}
\centering
%\begin{minipage}{60mm}
\caption{Pattern frequencies of figure rotation of triaxial Dehnen models with $\gamma=1$, $M_{bh}=0$}
\begin{tabular}{cccc}
\hline\\
$R_{\Omega}$  &  $\Omega_p$  & $\Omega_p^*$  & $T_p^* $ \\ 
 ($a$)  & (program units)  &(\kmskpc) & (yrs)\\
    \hline\\
%    \hline
 2  & 2.57$\times 10^{-1}$ & 33.8  &  $1.82\times10^8$\\
  5 & 7.45$\times 10^{-2}$ & 9.80  &  $6.28\times 10^8$\\  
 % 6 & 6.30$\times 10^{-2}$ & 8.29  &  $7.43\times 10^8$\\
10 & 2.87$\times 10^{-2}$ & 3.78  &  $1.63\times 10^9$\\  
20 & 1.09$\times 10^{-2}$ & 1.44  &  $4.29\times 10^9$\\
40 & 3.85$\times 10^{-3}$ & 0.51  &  $1.21\times 10^{10}$\\ 
60 & 2.12$\times 10^{-3}$ & 0.28  &  $2.21\times 10^{10}$\\
$\infty$ & 0.0                     & 0.0    &              $\infty$\\
\hline\\
%\hline
\multicolumn{4}{l}{{\it $^*$} For $a=5$~kpc, $M=5\times 10^{11}M_{\odot}$}
\label{tab:rotation}
\end{tabular} 
%\end{minipage}
\end{table}

\subsection{Orbit start spaces}
\label{sec:start}

Following \citet{merritt_fridman_96}, the model's mass distribution is
stratified into 20 ellipsoidal shells dividing the model into 21
sections of equal mass. In this paper, we focus mostly on orbits with
energy equal to the potential energy at the point the 8th shell
intersects the major axis of the model ($x = 1.6a$). (Note that shell
10 corresponds to the half-mass radius of the model). In the rotating
potential, orbits were launched from the equi-effective-potential
surface which is analogous to launching all orbits in a non-rotating
model from an equipotential surface. Thus all orbits launched from a
given shell have the same Jacobi integral ($E_J = E -
\frac{1}{2}|{\bf\Omega_p\times r}|^2$), where $E$ is the total energy
of an orbit. We study orbits on 6 energy shells for one pattern
frequency but defer a full investigation of orbits in self-consistent
potentials to a future paper \citep{valluri_11}.

Initial conditions for the orbits were selected in two different ways
such that orbits from all four major families (the box orbits and
three families of tube orbits) were represented
\citep{dezeeuw_85}. Although models are rotating, the orbits were
launched, integrated, examined, and classified only in the frame that
is corotating with the figure, allowing for a more direct comparison
with the stationary models that have been studied in the past.

In a stationary potential, box orbits are characterized by a
stationary point on the equipotential surface. In a rotating potential
there are no true box orbits (i.e., orbits with stationary points in an
inertial frame). However, orbits that are characterized by a
stationary point on the effective potential surface (surface of
constant Jacobi integral) look and behave much like box orbits in the
frame that is corotating with the figure. Accordingly, we launch
orbits at zero velocity on a regular grid on one octant of the
effective potential surface to obtain box-like orbits. In a
non-rotating triaxial model, a tube orbit is characterized by a finite
angular momentum (about either the long or short axis) which
oscillates between two values of the same sign. Thus the magnitude of
the angular momentum is not a conserved quantity, but the sign of the
angular momentum of a tube orbit remains constant \footnote{We refer
  to tube orbits as ``anti-clockwise'' or ``clockwise'' depending on
  whether their time average angular momentum vector ($J_x$ for
  $x$-tubes and $J_z$ for $z$-tubes) is positive or negative,
  respectively (in a right-handed coordinate system)}.  Consequently,
tube orbits avoid the center.

\begin{figure}
%%% FIGURE 1%%%
\begin{center}
\includegraphics[scale=.45,angle=0]{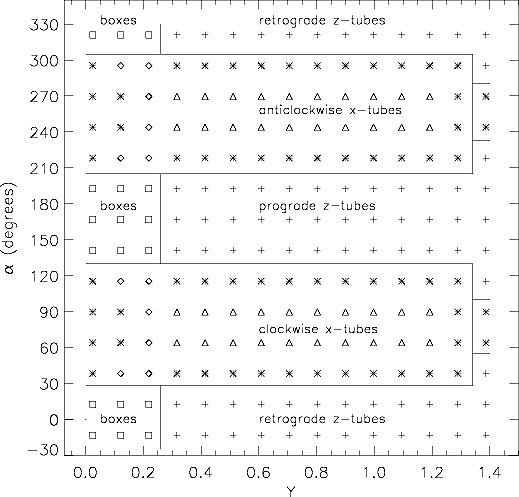}
\caption{Location of different orbit families on the $Y$-$\alpha$ start
  space in a stationary triaxial Dehnen model with $c/a = 0.5,
  T=0.58$, $\gamma = 0$, and $M_{bh}=0$. The lines mark approximate
  boundaries between major families. Squares ($\Box$) denote box
  orbits, asterisks ($\ast$) denote chaotic orbits, diamonds
  ($\diamond$) indicate inner long-axis ($x$) tubes, triangles
  ($\triangle$) denote outer long-axis tubes, and plus signs ($+$)
  indicate $z$-tubes.
\label{fig:yalpha_start}}
\end{center}
%%% FIGURE 1%%%
\end{figure}

For exploring the phase space structure of a stationary triaxial model
it is customary \citep{schwarzschild_93, merritt_fridman_96,
  vandenbosch_etal_08} to launch orbits uniformly from the ``$x$-$z$
start space'' i.e., from one quadrant of the $x$-$z$ plane with $v_x =
v_z = 0$ with $v_y >0$ determined by the energy of the equipotential
surface, this is adequate for stationary models since the properties
of orbits with $v_y <0$ are simply obtained by relying on the
symmetries of the model. However, in a triaxial model with figure
rotation about the short axis, orbits that are launched with $v_y>0$
in the $x$-$z$ start space are different from those with $v_y <0$. Since
orbits of all major families intersect the intermediate $y$-axis we
use instead the ``$Y$-$\alpha$ start space'' \citep{schwarzschild_82}
where all four families of orbits are launched from the $y$-axis of
the model with $v_y=0$ and at an angle $\alpha$ between the starting
velocity vector (perpendicular to the $y$-axis) and the $x$-$y$ plane. The initial conditions for an orbit are given by
\begin{align*}
x       &=  0                    & y      &=  Y  & z      &=  0\\
v_x   &=  V \cos \alpha & v_y  & =  0  & v_z  &= V \sin \alpha, 
\end{align*}
where $V$ is the magnitude of the total velocity of the orbit at that
position determined from potential energy at the starting point and
the Jacobi integral of all orbits on that surface.  By allowing
$0^\circ \le \alpha \le 360^\circ$ we obtain tube orbits with both
clockwise and anti-clockwise motions.  Figure~\ref{fig:yalpha_start}
shows the $Y$-$\alpha$ start space with 225 different initial conditions
marked by their orbital types. (We plot $-30<\alpha< 330$ to
illustrate that the retrograde $z$-tubes wrap around the top and
bottom boundaries.) The maximum value of $Y$ is determined by the
intersection of the equipotential surface with the $y$-axis.

The orbital types were determined visually by plotting three different
projections in coordinate space ($x-y$, $y-z$, and $x-z$) and by
examining their time-averaged normalized angular momenta about the
three principal axes (i.e., $\langle{J_x/|J_x|}\rangle$,
$\langle{J_y/|J_y|}\rangle$, and $\langle{J_z/|J_z|}\rangle$). For a box
orbit (or strongly chaotic orbit), all three components of normalized
angular momentum are approximately zero. For $x$-axis tubes and
$z$-axis tubes, the absolute values of the $x$ and $z$ components are equal
to unity (within 0.005\%). Note that a weakly chaotic orbit associated
with a tube family will have $\langle{J_x/|J_x|}\rangle \sim 1$ or
$\langle{J_z/|J_z|}\rangle \sim 1$ over a finite number of orbital
periods. In a three dimensional potential, Arnold diffusion would
cause such an orbit to eventually ergodically fill the entire energy
surface available to it and would therefore not be identifiable as a
tube orbit \citep{lichtenberg_lieberman_92}. Hereafter, we refer to
such orbits as ``tube-like''. Note that this criterion for classifying
orbits into three major families relies purely on the relative magnitudes
of the three components of the time-averaged angular momenta but does not
indicate that these orbits are regular.

Figure~\ref{fig:yalpha_start} shows how orbits of different types are
distributed on the $Y$-$\alpha$ start space. Squares indicate box
orbits, diamonds indicate inner long-axis tubes, triangles indicate
outer long-axis tubes, plus signs indicate short-axis tubes while an
asterisk indicates an orbit identified as stochastic by its high
diffusion rate (see Section~\ref{sec:chaotic}). Note that since the
triaxial figure rotates in the anti-clockwise direction about the $z$-axis, we refer to anti-clockwise $z$-tubes as ``prograde'' and their
clockwise counterparts as ``retrograde'' (even when a specific model
is stationary). We will see in
Figure~\ref{fig:yalpha_fmap_diff_gamma_0} that stochastic orbits are
frequently found along the boundaries (or separatrix layers) between
regions occupied by different orbit families. The general location of
the different regular orbit families in this start space will remain
roughly the same as the potential is perturbed with non-zero cusp
slope $\gamma$, the addition of a central point mass, or as the
pattern frequency, orbital energy, or the shape of the figure are
altered.

\subsection{Numerical Analysis of Orbital Frequencies}
\label{sec:chaotic}

%%%%%%%%%%%START OF CHANGES%%%%%%%%%%%%%%

In fully integrable potentials like the St\"ackel potentials,
all orbits are confined to $N$-dimensional tori, with $N$
the number of degrees of freedom (dof), i.e., the number of
spatial dimensions.
Motion on the torus is defined in terms of the fundamental 
frequencies $\omega_i$, $i=1,...,N$, 
the rates of change of the corresponding angle variables.
Expressed in terms of Cartesian coordinates, the motion 
is quasi-periodic, e.g.,
\begin{equation}
x(t) = \sum_{k=1}^{\infty} A_{k} e^{i\nu_kt},
\label{eq:fourierseries}
\end{equation}
where the $\nu_k$'s are linear combinations, with
integer coefficients, of the three fundamental frequencies:
\begin{equation}
\nu_k = n_{1,k}\omega_1+n_{2,k}\omega_2+n_{3,k}\omega_3
\end{equation}
 and the $A_k$ are the corresponding amplitudes. 
The same is true for regular orbits in arbitrary potentials,
i.e., for orbits that respect at least three isolating integrals 
of the motion \citep[e.g.,][]{lichtenberg_lieberman_92}.

When a regular orbit is
followed for many ($\sim 100$) dynamical times, a Fourier transform
of the trajectory yields a spectrum with discrete peaks. The
locations of the peaks in the spectrum correspond to the frequencies
$\nu_k$ in Equation~(\ref{eq:fourierseries}) and can be used to compute the three fundamental frequencies and the integer coefficients $(n_{1,k},n_{2,k},n_{3,k})$ 
that correspond to each peak
%\citep{boozer_82, kuo-petravic_etal_83, binney_spergel_82,
\citep[e.g.,][]{binney_spergel_82}.

\citet{laskar_90} developed a fast and accurate numerical technique
(``Numerical Analysis of Fundamental Frequencies,'' hereafter NAFF) to
decompose a complex time series representation of the phase space
trajectory of the form $x(t) + i v_x(t)$.  Our own implementation of Laskar's 
algorithm uses integer programming to obtain the fundamental frequencies 
from the spectrum \citep[][hereafter VM98]{valluri_merritt_98}. 
We integrated orbits using an explicit Runge Kutta
integrator (DOP853) of order eight (5,3) due to Dormand \& Prince with
stepsize control and dense output by Harier and Wanner
\citep{harier_etal_93}. The orbit integration, orbit classification,
and frequency analysis were all carried out in the frame that is
corotating with the pattern frequency of the figure.

Even for regular orbits, the character of the motion depends critically on whether the 
three fundamental frequencies are independent or whether they satisfy one or more 
nontrivial linear relations of the form
\begin{equation}
\ell\omega_1+m\omega_2+n\omega_3= 0
\label{eq:res}
\end{equation}
where $(\ell,m,n)$ are integers, not all of which are zero.
Generally there exists no relation like Equation (\ref{eq:res});
the frequencies are incommensurable,
and the trajectory fills its torus uniformly and densely 
in a time-averaged sense.
When one or more resonance relations are satisfied, however, 
the trajectory is restricted to a phase space region of 
lower dimensionality than $N$.

In 2 dof systems, a resonance implies a closed, or periodic, orbit,
e.g., a ``boxlet'' \citep{miralda_escude_schwarzschild_89}.
In three dimensions, a single resonance relation like 
Equation (\ref{eq:res}) does not imply that an orbit will be closed; 
rather, it restricts the orbit to a space of dimension two
\citep{merritt_valluri_99}.
An orbit satisfying one such relation is
``thin,'' confined for all time to a (possibly self-intersecting)
membrane.
In order for an orbit in a 3 dof system to be
closed, it must satisfy two such independent relations; 
such orbits are likely to be much rarer than thin orbits.
In what follows, we will use the term ``resonant'' to refer
both to thin orbits (satisfying one relation like Equation~\ref{eq:res})
and to closed orbits (satisfying two such relations), e.g.,
boxlets.
Stable resonances of both sorts generate new families of regular orbits 
whose shape mimics that of the parent orbit.
Unstable resonant tori are typically associated with 
a breakdown of integrability and with chaos.

%%%%%%%%%%%END OF CHANGES%%%%%%%%%%%%%

\begin{figure*}
\centering
%\begin{tabular}{cc}
\includegraphics[scale=.45,angle=90]{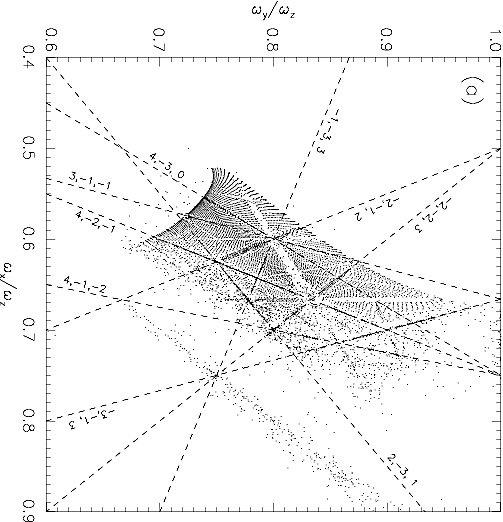}
\hfill
\includegraphics[scale=.45,angle=90]{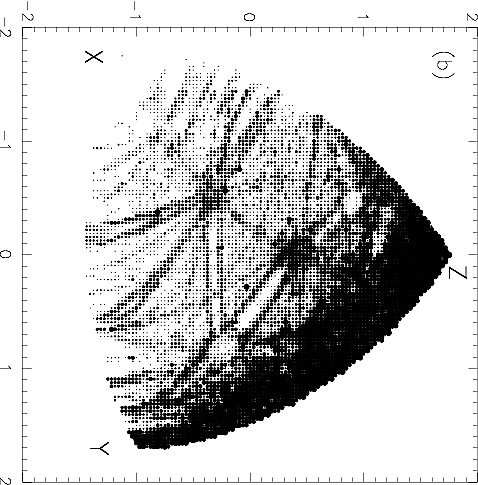} 
%\includegraphics[scale=.45,angle=90]{figure2b.jpg} 
%\end{tabular}
\caption{Two representations of phase space for 9408 box orbits
  launched from shell 8 in a non-rotating triaxial Dehnen model
  ($c/a=0.5$, $T=0.58$, $\gamma = 0$, and $M_{bh}=0$). (a) A frequency
  map: ${\omega_x}/{\omega_z}$ versus ${\omega_y}/{\omega_z}$ for the
  fundamental frequencies given by the NAFF algorithm.  Dashed lines
  mark various important resonances labeled by their integers $(\ell, m,$ and $n)$. (b) Diffusion map: gray scale corresponds to the
  diffusion parameter ($\log(\Delta f)$) of orbits at various initial
  positions on one octant of the equipotential surface (the
  ``stationary (box orbit) start space''). Dark regions on the map correspond to
  chaotic orbits and white regions correspond to regular orbits. The
  labels "X", "Y", and "Z" mark the intersections of the equipotential
  surface with the three principal axes.
\label{fig:box_start_diff}}
\end{figure*}

\begin{figure*}
\centering
\begin{tabular}{cc}%
\includegraphics[scale=.45,angle=0]{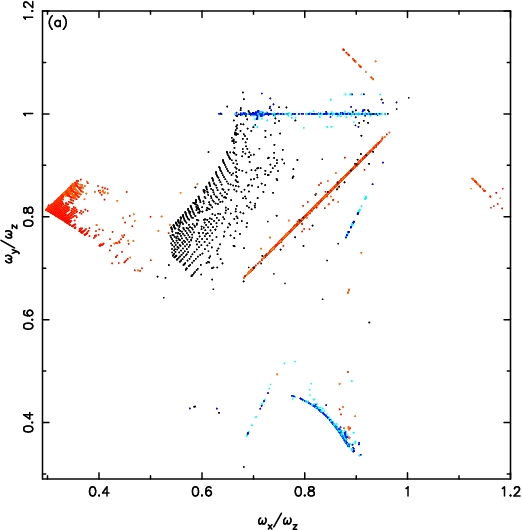}
\hfill
\includegraphics[scale=.45,angle=0]{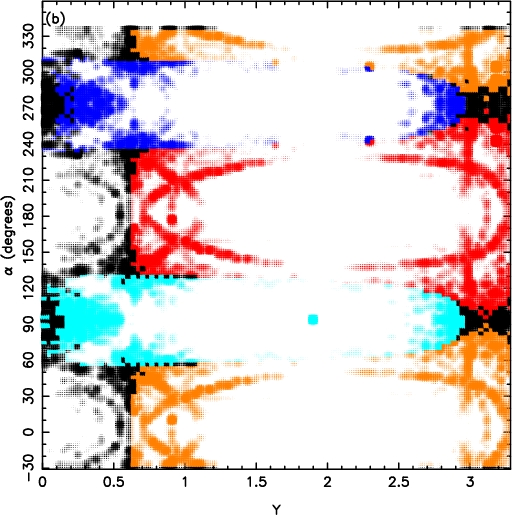}
\end{tabular}
\caption{(a) Frequency map and (b) diffusion map for 10,000
  orbits in the ``$Y$-$\alpha$ start space'' for a model with
  $c/a=0.5,T=0.5,$ and $\gamma=0$. Orbit families are color coded so that box
  orbits are black, anti-clockwise $x$-tubes are blue, clockwise
  $x$-tubes are cyan, clockwise (retrograde) $z$-tubes are ochre, and
  anti-clockwise (prograde) $z$-tubes are red. In the diffusion map
  only the chaotic orbits appear colored since the intensity of the
  color depends on the orbital diffusion parameter.
\label{fig:yalpha_fmap_diff_gamma_0}}.  
\end{figure*}

Once the fundamental frequencies and their amplitudes are obtained, it
is possible to obtain two complementary representations of phase space
at a given energy. A {\em frequency map} is obtained by plotting
ratios of pairs of frequencies (e.g., $\omega_x/\omega_z$
versus $\omega_y/\omega_z$) for many thousands of orbits in the
potential.  Such a representation of phase space
(e.g., Figure~\ref{fig:yalpha_fmap_diff_gamma_0}(a) and
Figure~\ref{fig:box_start_diff}(a)) is particularly useful for identifying
the most important orbital resonances. Stable resonances appear as filled lines with an increased density of orbits clustered along them. This is because these orbits have been trapped by the stable resonance.  In contrast, unstable resonances appear as ``blank'' or depopulated lines and are associated with stochastic orbits.

\citet{laskar_90} demonstrated that since most orbits in realistic
galactic potentials are only weakly chaotic, they lie close to regular
orbits in phase space mimicking their regular behavior over finite
time intervals.  Consequently, frequency analysis can be used to
distinguish between regular and chaotic orbits.  The frequency drift
can give a measure of the degree of chaos in an orbit. The frequency
drift of an orbit can be determined from the change in its fundamental
frequencies measured over two consecutive time intervals.
 
In applying Laskar's formalism, we integrated each orbit for $100 T_D$, 
where $T_D$ is the period of
the long-axis orbit of the same energy in the stationary model. The
time interval was divided into two equal segments labeled $t_1$ and
$t_2$ and three fundamental frequencies $\omega_x(t_1), \omega_y(t_1),
\omega_z(t_1)$ and $\omega_x(t_2), \omega_y(t_2), \omega_z(t_2)$ were
computed in each time segment. The ``frequency drift'' or ``diffusion
parameter'' in each frequency component is given by
\begin{subequations}
\begin{align}
%\begin{eqnarray}
 \log(\Delta f_x) & = \log{\left\vert{\frac{\omega_x(t_1)-\omega_x(t_2)}{\omega_x(t_1)}}\right\vert},\\
 \log(\Delta f_y) & =  \log{\left\vert{\frac{\omega_y(t_1)-\omega_y(t_2)}{\omega_y(t_1)}}\right\vert},\\
\log(\Delta f_z) & = \log{\left\vert{\frac{\omega_z(t_1)-\omega_z(t_2)}{\omega_z(t_1)}}\right\vert}.
\label{eq:diffusion}
\end{align}
%\end{eqnarray}
\end{subequations}
We define the diffusion parameter $\log(\Delta f)$ to be the value
associated with the largest of the three amplitudes $A_x, A_y,$ and $A_z$. The larger the value of the diffusion parameter the more chaotic
the orbit. Our second representation of phase space, a {\em diffusion
  map}, is obtained by plotting the initial conditions of many
thousands of orbits and adding a gray scale (or color intensity scale)
corresponding to the diffusion parameter.  In a diffusion map
(e.g., Figure~\ref{fig:box_start_diff}(b) and
Figure~\ref{fig:yalpha_fmap_diff_gamma_0}(b)) the gray scale (or intensity
of the color) corresponds to $\log(\Delta f)$ such that regions of
initial condition space occupied by regular orbits are white and those
occupied by chaotic orbits are dark.

Several properties of the phase space can be inferred from frequency
and diffusion maps.  We begin with a discussion of phase space maps
for the stationary (here after "box orbit") start space for a baseline
model, namely a triaxial Dehnen model with $c/a=0.5, T=0.58, \gamma=0,$ and
$M_{bh}=0$. The frequency map in Figure~\ref{fig:box_start_diff}(a)
shows a plot of the ratios of fundamental frequencies
${\omega_x}/{\omega_z}$ versus ${\omega_y}/{\omega_z}$ for each of $9408$
box orbits dropped with zero velocity from the equipotential surface
corresponding to the 8th shell.

Note that although we focus primarily on the behavior of orbits
launched from shell 8th, the behavior of an orbit in response to
figure rotation depends both on the (radial) energy shell from which it
is launched as well as the co-rotation radius. However,
since the density profile has an approximately power-law profile
outside the inner cusp, the behavior of orbits depends primarily on
$R_\Omega/a_{i}$, where $a_{i}$ is the semi-major axis of the $i$th
shell. Consequently, the behavior of orbits launched from an outer
shell at a moderate pattern speed can resemble that of orbits launched
at an inner radius and a faster pattern speed. Consequently, we will
frequently give values of $R_\Omega/a_{i}$ as well as $R_\Omega$.

%%%%%%%%%%%START OF CHANGES%%%%%%%%%%%%%%%%
%The frequency map shows that most points at the bottom left corner of
%the plot lie on a fairly regular grid which reflects the regular
%distribution of initial conditions.  Away from this region the regular
%grid like structure is disrupted by the appearance of dark lines
%correspond to stable resonances which have trapped a significant
%number of orbits. All the orbits lying along such dark lines obey a
%single resonant condition such as: $l \omega_x + m\omega_y + n
%\omega_z = 0$.  All orbits that satisfying a resonance condition are
%``thin orbits'' since they cover the surface of a two dimensional
%surface in phase space \citep{merritt_valluri_99}.\footnote{In
%  contrast non resonant regular orbits in a 3-dimensional potential
%  conserve three integrals of motion and therefore densely fill a
%  three-dimensional volume in configuration phase (BT08).}  In the
%plot, several such resonances are highlighted with dashed lines and
%are labeled with their integers ($l, m, n$). The strength of a given
%resonance (or its ability to trap orbits) can be determined by the
%number of orbits that are associated with it, or in the case of
%unstable resonances, the width of the region of the frequency map that
%has been evacuated.

The frequency map shows that most points at the bottom left corner of
the plot lie on a fairly regular grid which reflects the regular
distribution of initial conditions.  Away from this region the regular,
grid-like structure is disrupted by the appearance of dark lines
corresponding to stable resonances and their associated regular orbits. 
All the orbits lying along such lines obey a
single resonant condition like that of Equation~(\ref{eq:res}).
In the plot, several such resonances are highlighted with dashed lines and
are labeled with their defining integers ($\ell, m,$ and $n$).

%%%%%%%%Figure 2%%%%%%%%%
%%%%%%%%% Figure 2 %%%%%%%%%%%%%

%%%%%%%%% Figure 3 %%%%%%%%%%%
%%%%%%%%% Figure 3 %%%%%%%%%%%

Figure~\ref{fig:yalpha_fmap_diff_gamma_0}(a) shows a frequency
map for 10,000 orbits initialized on the $Y-\alpha$ start space with
the energy corresponding to the $x$-axial orbit on shell 8 in this
model. In this frequency map, all four major orbit families are
represented: box orbits are black, anti-clockwise $x$-tubes are blue,
clockwise $x$-tubes are cyan, clockwise $z$-tubes are ochre, and
anti-clockwise $z$-tubes are red.  Note that a significant fraction of
the $z$-tubes (red and ochre points) and $x$-tubes (blue and cyan
points) appear clustered along straight lines.  These are orbits that
lie close to the thin-shell tube orbit parent of the family and appear
as ``resonance lines'' in the frequency map although they are not
traditionally viewed as resonances.  To accurately represent the
resonances in the two tube orbit families it is essential to obtain
the fundamental frequencies in cylindrical coordinates about the
appropriate symmetry axis.  VM98 showed that such a representation
reveals several unstable tube orbit resonances. These can be seen as
dark bands in Figure~\ref{fig:yalpha_fmap_diff_gamma_0}(b).

Figure~\ref{fig:box_start_diff}(b) is a diffusion map for 9408
box orbits launched with zero initial velocity from the equipotential
surface (at shell 8) of a stationary maximally triaxial Dehnen model
with $\gamma=0$. The large swath of regular orbits at the bottom left
of the diffusion plot consists of regular box orbits that originate
close to the $x$-axis. In addition, several white bands mark regular
islands of resonant boxlet families. Each white band corresponds to a
dark resonance line in the frequency map in
Figure~\ref{fig:box_start_diff}(a). Most white bands are flanked
by narrow dark regions (occupied by stochastic orbits) which lie along
the ``separatrix'' (transition layer) between different orbit families.
Stochastic orbits (dark regions) are also seen at the intersections of
stable resonances (white bands), when the parent periodic orbit is
unstable. The prominent dark band of chaotic orbits that runs along
the right edge of the diffusion map corresponds to orbits originating
close to the $y$-$z$ plane. Orbits in this region are chaotic due to the
well-known instabilities of the $y$-axial and $z$-axial orbits
\citep{heiligman_schwarzschild_79,heisler_etal_82,dezeeuw_merritt_83,adams_etal_07}
as well as due to the instability of orbits confined in the $y$-$z$
plane to perturbations perpendicular to this plane
\citep{adams_etal_07}. This region will be seen to expand with
increasing figure rotation and is referred to hereafter as the ``$y$-$z$
instability band''. Most of these features were described in detail in
previous papers \citep{valluri_merritt_98, merritt_valluri_99}.

Figure~\ref{fig:yalpha_fmap_diff_gamma_0}(b) shows a diffusion map for
10,000 orbits launched from the $Y$-$\alpha$ start space. The orbits were
classified into three different orbit families based on their time-averaged normalized angular momentum values. The boxes are colored
black, anti-clockwise $x$-tubes are blue, clockwise $x$-tubes are
cyan, clockwise (retrograde) $z$-tubes are ochre, and anti-clockwise
(prograde) $z$-tubes are red.  Note that regions of more intense
color (occupied by stochastic orbits) generally appear at the
separatrix between the major orbit families. Within the regions
occupied by $x$-axis tubes (blue and cyan regions), the transition
between the inner and outer $x$-axis tubes is also marked by a weakly
stochastic layer. Thick stochastic layers separate the $z$-tubes from
the box orbits that lie at values of $Y <0.5$. In addition, stochastic
orbits are found at $\alpha \sim 90^\circ\pm30^\circ$ and $\alpha \sim
270^\circ\pm30^\circ$ mostly at very small $Y$ values (these orbits
are launched nearly along the $z$ axis which is known to be unstable).

The main features identified on the frequency and diffusion maps
above, for the $\gamma=0$ and $M_{bh}=0$ case, are meant to give readers
insight into how they should interpret the results for models with
figure rotation.

\section{Results}
\label{sec:results}

We now discuss the effects of increasing the pattern frequency of
figure rotation on box-like and tube-like orbits using the phase-space
mapping techniques discussed in the previous section.

\subsection{Effects of slow/Moderate pattern frequency on boxlike orbits}
\label{sec:box}

Box orbits were integrated, from start spaces such as the one in
Figure~\ref{fig:box_start_diff}(b), for various model parameters
and increasing pattern speeds.  We generally present results for a
default model which is  close to maximally triaxial, with $c/a=0.5,
T= 0.58$ and central density cusp of $\gamma=1$. We also present
results for models with a central point mass representing a central
SMBH with $M_{bh} = 0.001$ (0.1\% of the galaxy
mass). (Models with $\gamma=2$ were studied but are not presented
since their dependence on $R_\Omega$ is qualitatively similar although
they have a higher fraction of chaotic orbits, VM98. Higher black hole mass fractions were previously studied by VM98 and were found to induce more chaos.) In each figure,
we show phase space maps for a non-rotating model ($R_\Omega=\infty$)
and three additional values of the corotation radius ($R_\Omega = 40,
10,$ and $5$). In all cases orbits were launched with energy $E$ equal to
that of the $x$-axial orbit started from the 8th shell. Therefore, for an orbit launched from the 8th shell, these pattern speeds correspond to $R_\Omega/a_i = 24.8, 6.2,$ and $3.1$.

%%%%%%%% Figure 4 %%%%%%%%%
\begin{figure*}
\centering
\begin{tabular}{cc}
\includegraphics[scale=.4,trim= 4.3mm 4.3mm 0mm 0mm, clip,angle=0]{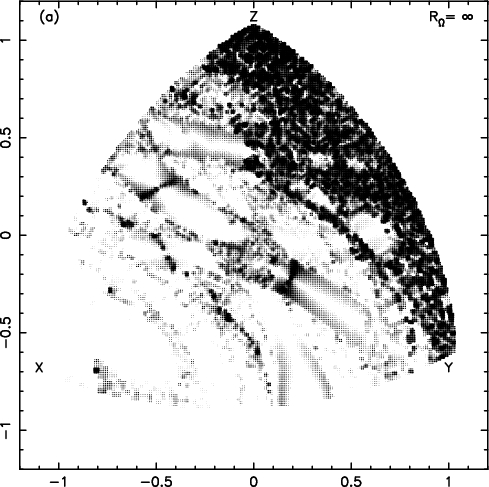}
\includegraphics[scale=.4,trim= 4.3mm 4.3mm 0mm 0mm, clip,angle=0]{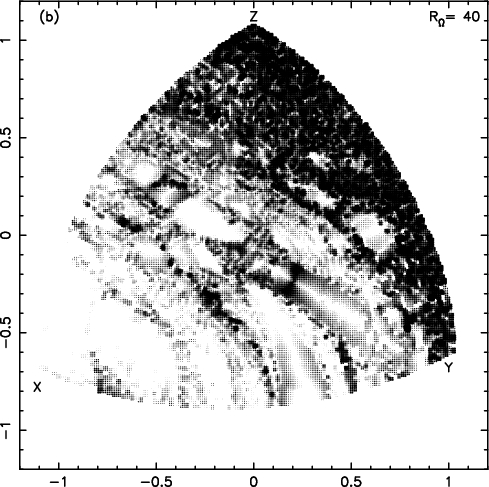}\\ 
\includegraphics[scale=.4,trim= 4.3mm 4.3mm 0mm 0mm, clip,angle=0]{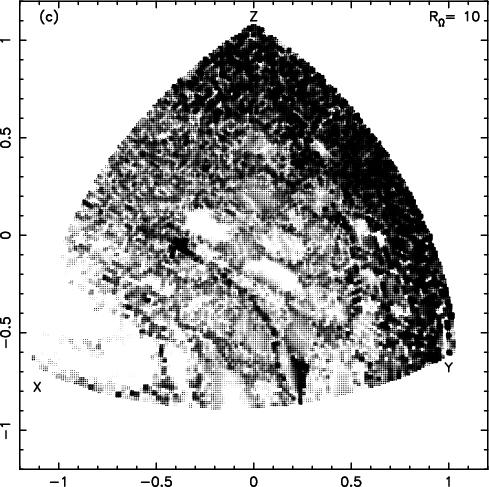}
\includegraphics[scale=.4,trim= 4.3mm 4.3mm 0mm 0mm, clip,angle=0]{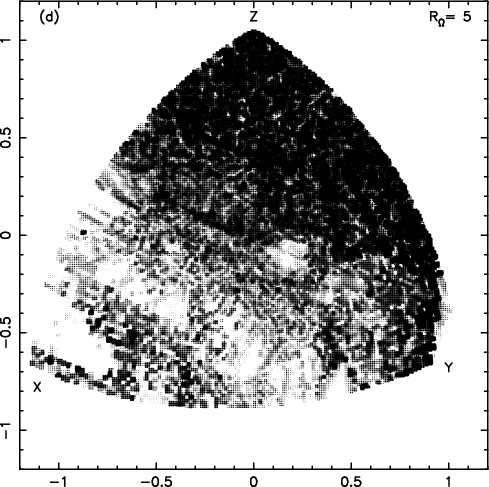}\\
\end{tabular}
\caption{Diffusion maps of orbits launched from the box orbit
  start space at 8th energy shell in models with $\gamma = 1$ and
  $M_{bh} = 0$ and varying pattern frequencies as indicated by labels.
  The value of $R_\Omega = {\rm \infty}$ refers to a model with no
  rotation, while $R_\Omega = 40, 10,$ and $5$ refer to the corotation
  radii of models with increasing pattern frequencies. These pattern speeds correspond to $R_\Omega/a_i = 24.8, 6.2,$ and $3.1$.
\label{fig:box_diff_nobh}}
\end{figure*}
%%%%%%%% Figure 4 %%%%%%%%%

Figure~\ref{fig:box_diff_nobh} shows phase space diffusion maps for
box orbits in our default model with no central
point mass. The non-rotating model (panel (a)) is similar to
(Figure~\ref{fig:box_start_diff}(b)) but contains a somewhat larger
fraction of stochastic orbits.   Almost all the regular orbits (white regions on the map) are now associated with resonant islands containing boxlet
families. All orbits that qualitatively resemble box orbits in cored
potentials are mildly to strongly chaotic. 

As mentioned in Section~\ref{sec:intro}, \citet{gerhard_binney_85} first
proposed that the box orbits rendered chaotic due to scattering by a massive central point mass could be stabilized in a rotating frame
due to the envelope doubling effect of the Coriolis force. Since a box
orbit (in a stationary model) has no net angular momentum it
oscillates between stationary points reversing its sense of
progression around the center each time it reaches a turning point. In
a rotating frame this means that the path described during the
prograde segment of the orbit is not retraced during the retrograde
segment because the Coriolis force on the two segments differs
\citep{schwarzschild_82,dezeeuw_merritt_83}.

\begin{figure*}
\centering
\begin{tabular}{cc}
\includegraphics[scale=.4,trim= 4.3mm 4.3mm 0mm 0mm, clip,angle=0]{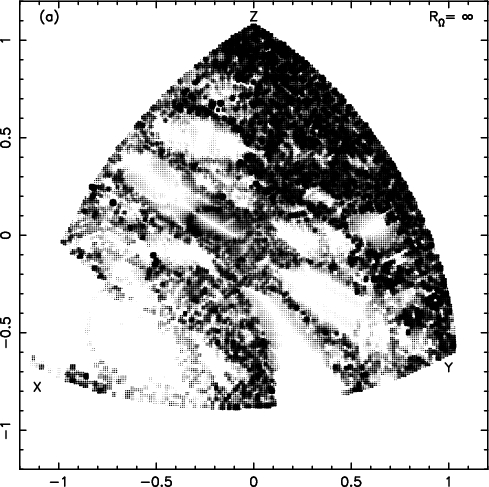}
\includegraphics[scale=.4,trim= 4.3mm 4.3mm 0mm 0mm, clip,angle=0]{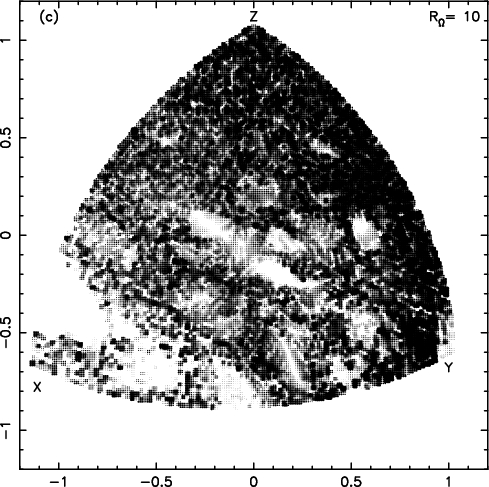}\\ 
%%%REDO FIGURE Boxx202_diff.ps (Figure 5c) to improve quality !!!
\includegraphics[scale=.4,trim= 4.3mm 4.3mm 0mm 0mm, clip,angle=0]{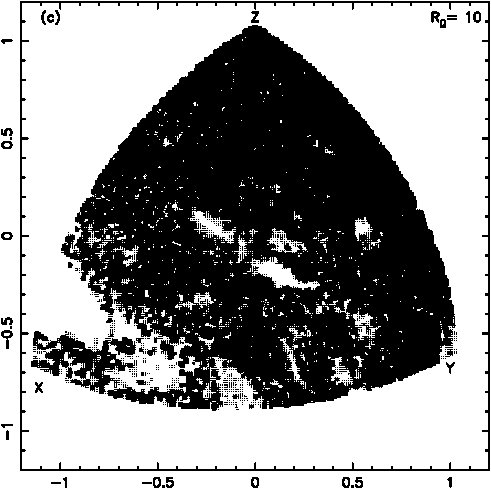}
\includegraphics[scale=.4,trim= 4.3mm 4.3mm 0mm 0mm, clip,angle=0]{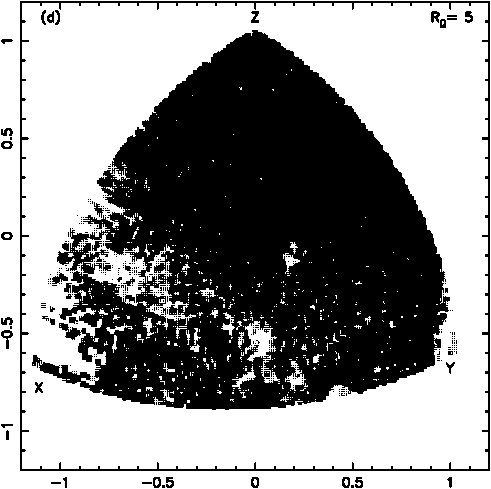}\\
\end{tabular}
\caption{Diffusion maps of orbits launched from the box orbit start space on the 8th energy shell of   a model with $c/a=0.5, T= 0.58, \gamma = 1$, and $M_{bh} = 0.001$. The value of $R_\Omega =
{\rm \infty}$ refers to a model with no rotation while $R_\Omega = 40,
10,$ and $5$ refer to the corotation radii of models with increasing pattern
frequencies. 
\label{fig:diff_wbh}}
\end{figure*}

We see in Figure~\ref{fig:box_diff_nobh} that this prediction does not
hold up in realistic triaxial galaxy models. In fact, as the pattern
frequencies of figure rotation increase ($R_\Omega$ decreases) the
area of the diffusion map occupied by regular orbits (white) decreases
and the resonant islands shrink until only a small fraction of orbits
remain regular at $R_\Omega=5$.  When a central point mass is added,
the increase in the fraction of chaotic orbits is even more
significant (Figure ~\ref{fig:diff_wbh}) and in addition the delineation
between the various resonant box orbit families is even less clear $-$ 
pointing to increased resonance overlap. We see from a comparison of the Figures~ 4 and 5 that high pattern speed increases the fraction of chaotic orbits more significantly than the presence of a central black hole.

%%%%%%%% Figure 5 %%%%%%%%%
%%%%%%%% Figure 5 %%%%%%%%%

We note in passing that a new regular orbit family appears as a white
region in the lower right hand corner of the diffusion maps in
Figures~\ref{fig:box_diff_nobh} and~\ref{fig:diff_wbh} for
$R_\Omega = 10, 5$. In the stationary model, box orbits originating
here are highly unstable since they lie close to the unstable
$y$-axis. Figure~\ref{fig:yaxis_ztube}(top row) shows two
projections of one such orbit in a model with $R_\Omega=5$.  Although
it is launched with no net angular momentum in the rotating frame,
this orbit is a short-axis tube. This tube orbit (like others in
triaxial potentials) does not conserve angular momentum, but its value
oscillates between two values (in this case zero and a large negative
value) indicating that its motion in the rotating frame is retrograde
to that of the figure. This new family of orbits lies close to the
$x$-$y$ plane. This family appears to arise when the pattern frequency
of the figure is high enough that Coriolis and centrifugal forces
cause it to loop around the center. Orbits in this region of start
space were chaotic in the stationary model and are indeed stabilized
by figure rotation as predicted by \citet{gerhard_binney_85}. However,
this stabilization results from a complete transformation of the
orbital character from box-like to tube-like, rather than due to a small
deflection of the chaotic box orbit around the destabilizing
center. We will see later (Section~\ref{sec:fast}) that as the corotation
radius moves inward, this region of stable tube-like orbits grows
steadily till most of the box start space is occupied by regular
orbits, similar to this one.

Figures~\ref{fig:box_diff_nobh}(a) and~\ref{fig:diff_wbh}(a) show that most (but not all) the
regular orbits in the stationary model, which appear as white regions
on the diffusion maps, are associated with resonant orbits.
There  are two reasons why the majority of box orbits (which are associated with two-dimensional resonance surfaces) are not stabilized by slow to moderate figure rotation as predicted by \citet{gerhard_binney_85}.

%%%%%%% Figure 6 %%%%%%%%%%%%%%%%%
 \begin{figure}
\centering
\begin{tabular}{cc}
\includegraphics[scale=.5]{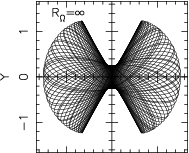}
\hspace{.5in}
\includegraphics[scale=.5]{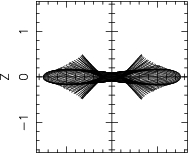}\\
\includegraphics[scale=.5]{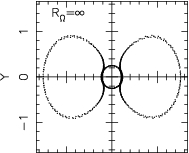}
\hspace{.5in}
\includegraphics[scale=.5]{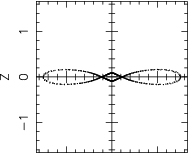}\\
\includegraphics[scale=.5]{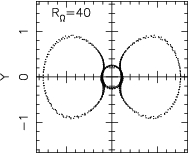}
\hspace{.5in}
\includegraphics[scale=.5]{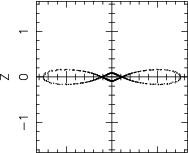}\\
\includegraphics[scale=.5]{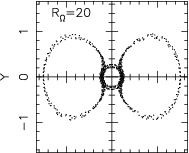}
\hspace{.5in}
\includegraphics[scale=.5]{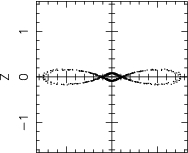}\\
\includegraphics[scale=.5]{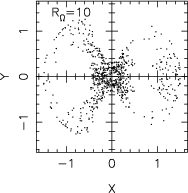}
\hspace{.5in}
\includegraphics[scale=.5]{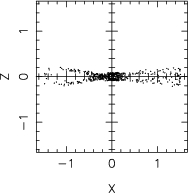}\\
%\includegraphics[scale=.5]{figure8a.jpg}
%\hspace{.5in}
%\includegraphics[scale=.5]{figure8b.jpg}\\
%\includegraphics[scale=.5]{figure8c.jpg}
%\hspace{.5in}
%\includegraphics[scale=.5]{figure8d.jpg}\\
%\includegraphics[scale=.5]{figure8e.jpg}
%\hspace{.5in}
%\includegraphics[scale=.5]{figure8f.jpg}\\
%\includegraphics[scale=.5]{figure8g.jpg}
%\hspace{.5in}
%\includegraphics[scale=.5]{figure8h.jpg}\\
%\includegraphics[scale=.5]{figure8i.jpg}
%\hspace{.5in}
%\includegraphics[scale=.5]{figure8j.jpg}\\
\end{tabular}
\caption{Resonant orbit associated with the (3,$-$1,$-$1)
  resonance.  The top row shows two Cartesian projections of the
  orbit. The next four rows show cross sections of the orbit with
  the $x$-$y$ plane (left) and with the $x$-$z$ plane (right) with
  increasing pattern frequencies ($R_\Omega = \infty, 40, 20,$ and $10$ which correspond to  $R_\Omega/a_i = 24.8, 12.4,$ and $6.2$).
\label{fig:resonant_orbits}}
\end{figure}
 %%%%%%%%% Figure 6 %%%%%%%%%%%%%%%%%

First, the increase in the fraction of chaotic orbits can be
attributed to the distinct configuration space structure of resonant
orbits. As described by \citet{merritt_valluri_99}, the parent of a
resonant orbit family satisfies a condition like $l \omega_x +
m\omega_y + n \omega_z = 0$.  This additional resonant condition
confines a resonant orbit to the surface of a two-dimensional torus in
three-dimensional phase space. In general, such orbits are not
periodic.  The more commonly known (but less numerous) near-periodic
versions of such resonances were identified by
\citet{miralda_escude_schwarzschild_89} and given names such as
``bananas'', ``pretzels'', ``fish'' etc. based on their projected
shapes. A non-periodic resonant orbit therefore, in comparison,
appears similar to a three-dimensional box orbit in projection but occupies only a
two-dimensional surface in configuration space.  Consequently, all
families of stable resonant orbits are centrophobic and avoid the
center of the potential due to their thin sheet-like structure. The
parent of the resonant family is surrounded by a region occupied by
nearly resonant orbits which are characterized by two of the same
frequencies as the resonant parent, but with an increasing third
frequency. As the third frequency increases from zero, the thickness
of the orbit increases and the nearest distance of approach to the
center of the potential (the pericenter distance) decreases. When the
orbit becomes thick enough to pass through the center it is
destabilized. The instability at the center is a consequence of the
divergent central density cusp or central black hole present in all
realistic galaxy models.

In the rotating frame of a slowly rotating model, the Coriolis force
can produce ``envelope doubling'' which broadens a nearly resonant box
orbit driving it into the divergent central cusp (or central black
hole), reducing the sizes of stable resonant islands, and eventually
destroying them. The effect of envelope doubling can be seen in
Figure~\ref{fig:resonant_orbits}, which shows a resonant boxlet orbit
associated with the (3,$-$1,$-$1) resonance. The top row shows two
Cartesian projections of the orbit, $x$-$y$ (left) and $x$-$z$
(right). In projection (top row) the orbit appears similar to three-dimensional box orbits. However, the second row shows intersections of the orbit with the $x$-$y$ plane (left) and the $x$-$z$ plane (right) shows its perfectly sheet-like structure with a clear ``hole'' in the center.  The next three rows
show cross sections of the same orbit in models with $R_\Omega=40, 20,$ and $10$. When $R_\Omega = 20$ (4th row) the cross section shows that the Coriolis force broadening has caused it to nearly fill the center in the $x$-$z$ projection.  When $R_\Omega = 10$ (5th row) the orbit is thick enough to pass through the center rendering it completely chaotic.
 
 While the majority of the regular orbits on the stationary start space in
 Figure~\ref{fig:box_diff_nobh} and Figure~\ref{fig:diff_wbh} are
 associated with resonances, a few regular box orbits are
 non-resonant. These too appear to be destabilized by figure
 rotation. Additionally, \citep{gerhard_binney_85} predicted that
 orbits that are chaotic in the stationary frame should become regular
 in the rotating frame, but this was not seen.
 
%%%%%%%% Figure 7 %%%%%%%%%
\begin{figure*}
\centering
\begin{tabular}{cc}
\includegraphics[scale=.4,angle=0]{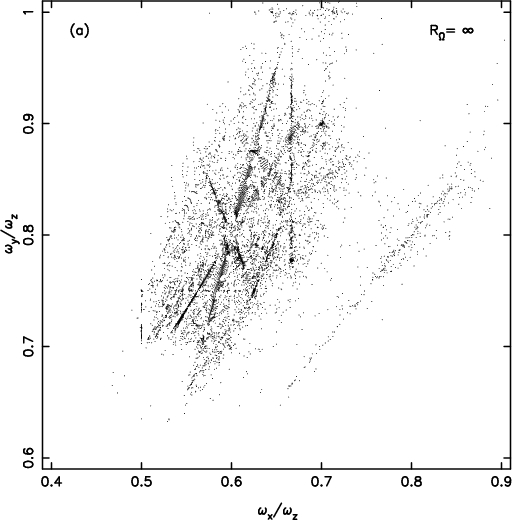}
\includegraphics[scale=.4,angle=0]{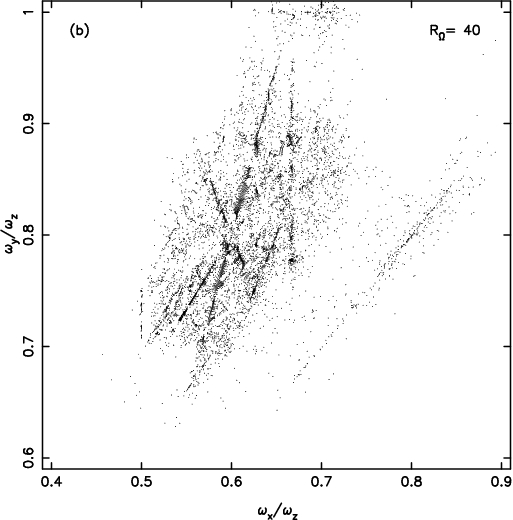}\\
\includegraphics[scale=.4,angle=0]{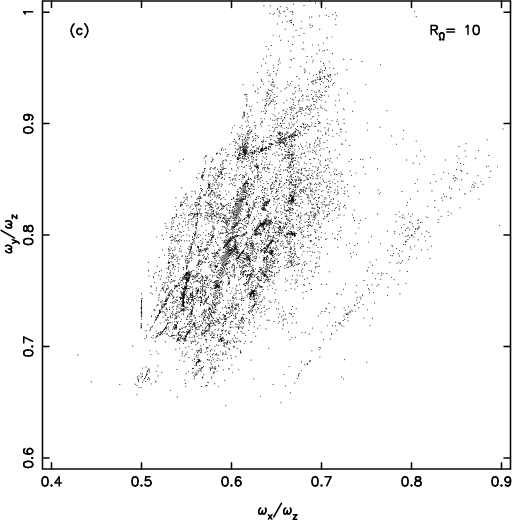}
\includegraphics[scale=.4,angle=0]{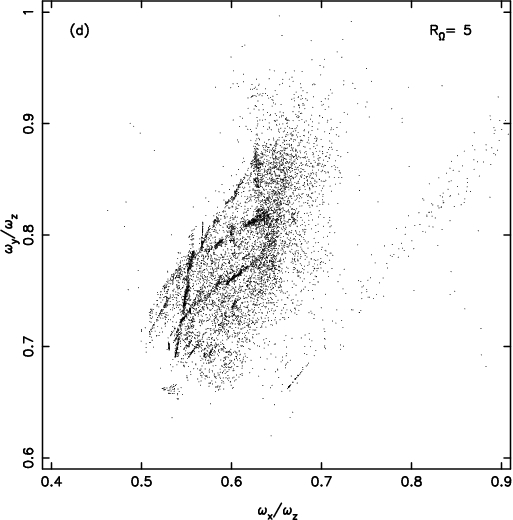}\\
\end{tabular}
\caption{Frequency maps for box orbit start space at shell 8 in the
  default model ($c/a=0.5$, $T=0.58$, $\gamma = 1$, and $M_{bh}=0$) for
  a non-rotating model ($R_\Omega=\infty$) and three rotating models
  ($R_\Omega= 40, 10,$ and $5$).
\label{fig:box_fmap_nobh}}
\end{figure*}
%%%%%%%% Figure 7 %%%%%%%%%

The second reason for the increase in the fraction of chaotic boxlike
orbits can be seen by plotting frequency maps. Frequency maps for the
default model ($c/a=0.5, T=0.58, \gamma= 1,$ and $M_{bh}=0$) show resonant
orbits clustered along resonance lines, and non-resonant regular
orbits and stochastic orbits scattered over the rest of the map
(Figure~\ref{fig:box_fmap_nobh}(a)).  The points scattered along a line with a slope of approximately unity on the right hand
side of the map correspond to chaotic box orbits which lie at the
separatrix boundary with a family of short-axis tubes with $\omega_x
\sim \omega_y$.

For the slowest rotation frequency ($R_{\Omega}=40$), we observe the
disappearance of some of the weakest resonances but the overall
structure of the frequency map remains intact. As the corotation
radius of the model decreases (higher rotation frequencies) the entire
frequency map appears to shrink with the boundaries moving toward the
bottom right toward the short-axis tube ``resonance''. This is because as the
pattern frequency of the model increases, each orbit experiences a
centrifugal force that changes both $\omega_x$ and $\omega_y$. For a
small fraction of orbits that lie close to the equatorial plane of the
model with large $y$ values, the change in orbital frequencies is
large enough to convert them to resonant $z$-tubes
(e.g., Figure~\ref{fig:yaxis_ztube} (top row)).

The frequency maps show the increased width of the stable resonance
lines (which occur due to the envelope doubling discussed above). In
addition a few new resonances appear as the pattern speed
increases. The shrinking of the frequency map, the broadening of
resonances, and the appearance of new resonances together
contribute to a significant increase in the overlap of
resonances. Resonance overlap is a well known cause of global chaos in
Hamiltonian systems \citep{chirikov_79} and may be thought of as
occurring when several different resonances compete to trap the same
orbit (BT08).  This implies that even if figure rotation could allow
chaotic box orbits to avoid the center as suggested by
\citet{gerhard_binney_85}, the compression of the range of frequencies
resulting from the modulation of orbit frequencies with the pattern
frequency results in such significant overlap in resonant orbits that
the effects of global chaos dominate the behavior of box orbits.

\subsection{Effect of moderate figure rotation on tubelike orbits}
\label{sec:tubes}

%%%%%%% Figure 8 %%%%%%%%%%
\begin{figure*}
\centering
\begin{tabular}{cc}
\includegraphics[scale=.4,angle=0]{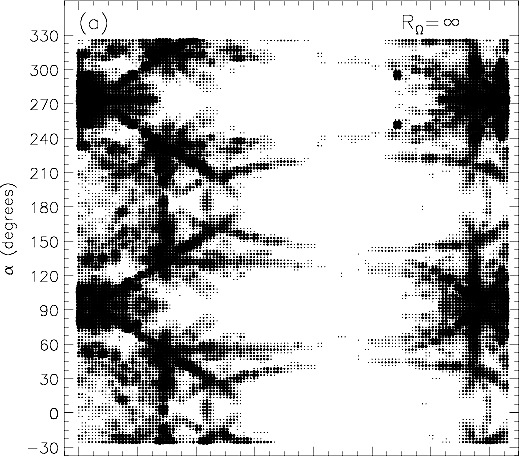}
\includegraphics[scale=.4,angle=0]{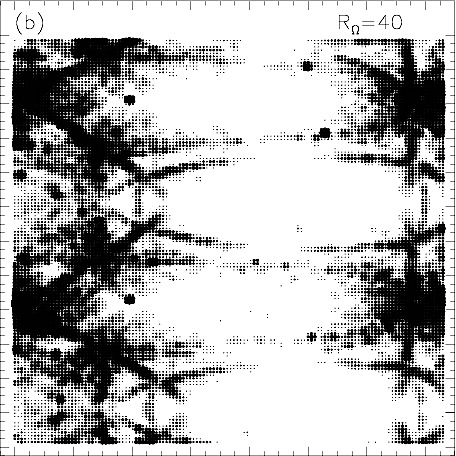}\\ 
\includegraphics[scale=.4,angle=0]{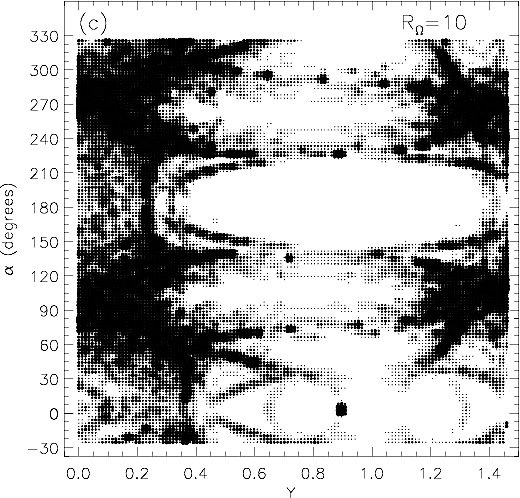}
\includegraphics[scale=.4,angle=0]{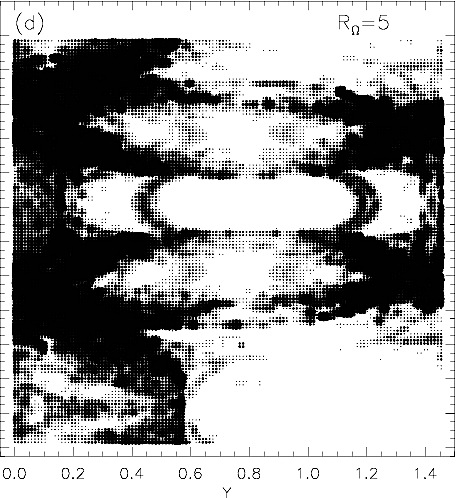}\\
\end{tabular}
\caption{Diffusion maps for orbits launched from the $Y$-$\alpha$ start
  space in a Dehnen triaxial model (with $c/a=0.5$, $T=0.58$, $\gamma
  = 1$, and $M_{bh}=0$). Panel (a) shows the model with no rotation
  ($R_\Omega=\infty$) while the other three panels show models with
  increasing pattern frequency ($R_\Omega= 40, 10,$ and $5$).
\label{fig:yalpha_diff_nobh}}
\end{figure*}
%%%%%%% Figure 8 %%%%%%%%%%%%

We now explore the effect of figure rotation on tube-like orbits in the
default model. The  diffusion maps in 
Figure~\ref{fig:yalpha_diff_nobh} show orbits launched from the
$Y$-$\alpha$ start space, with the gray scale
representing the diffusion parameters.  (Compare with
Figure~\ref{fig:yalpha_fmap_diff_gamma_0} for the locations of the major
orbit families.)

As the figure rotation of the model increases, the bi-symmetry about
$\alpha \sim 150^\circ$ is broken. For $R_\Omega=5$, the prograde
$z$-tubes shrink to occupy a region $\alpha \sim 180^\circ\pm20^\circ$
(i.e., close to the equatorial plane of the model) but now extend to
smaller $Y$ values than in the stationary model (panel (a)). In
contrast, the retrograde $z$-tubes now occupy a large range of
$\alpha$ values, but predominantly at large $Y$). As the co-rotation
radius decreases, the regions occupied by long-axis ($x$) -tubes also
shrink.  A greater portion of phase space at small values of $Y$
associated with box orbits becomes chaotic (consistent with what was
seen in Figure~\ref{fig:box_diff_nobh}).

\citet{binney_81} found  from stability of the analysis of
closed periodic (1:1) orbits in the equatorial plane that as figure
rotation increases, retrograde closed orbits become unstable in an annular region  (the ``Binney instability strip'') where the perpendicular frequency is resonant with the epicyclic frequency. Based on this finding,\citet{schwarzschild_82} erroneously concluded that retrograde tube orbits would become increasingly unstable with increasing figure rotation and consequently chose to largely omit them from the self-consistent models he constructed.

The decrease in the overall fraction of initial condition space occupied by the prograde $z$-tubes and the corresponding growth of the retrograde $z$-tubes appears to be in conflict with this expectation. However, we do find that at the highest pattern frequency (panel (d)), prograde $z$-tubes close to the equatorial
plane ($\alpha \sim180^\circ$) are stable over a wider range
of $Y$ values than the retrograde variety ($\alpha \sim0^\circ$). 

In the stationary model (panel (a)) the $x$-tubes are found
symmetrically about launch angles of $\alpha \sim 90^\circ\pm30^\circ,
270^\circ\pm30^\circ$ (orbits are launched perpendicular to the
$x$-$y$ plane). However, as the pattern frequency of figure rotation
increases the portion of start space occupied by $x$-tubes decreases
and orbits with $\alpha \lesssim 90$ and $\alpha \gtrsim 270$ are
replaced by retrograde $z$-tubes (at $Y\gtrsim 0.6$) and box (and
chaotic) orbits (at $Y \lesssim 0.6$). Thus, in the more rapidly
rotating models any orbit launched vertically (with no angular
momentum about the $z$-axis) or with velocity opposite to the
direction of figure rotation will ``fall behind'' the model. The orbit
can become a retrograde $z$-axis tube (in the frame corotating with
the figure), a box orbit, or a chaotic orbit. (Note that although some
orbits are ``retrograde'' in the corotating frame, they could be
rotating prograde when viewed from an inertial frame.)

\begin{figure*}
\centering
\begin{tabular}{cc}
%\includegraphics[scale=.5]{figure10a.ps}
%\hspace{.3in}
%\includegraphics[scale=.5]{figure10b.ps}
%\hspace{.7in}
%\includegraphics[scale=.5]{figure10k.ps}
%\hspace{.3in}
%\includegraphics[scale=.5]{figure10l.ps}\\ 
%\includegraphics[scale=.5]{figure10c.ps}
%\hspace{.3in}
%\includegraphics[scale=.5]{figure10d.ps}
%\hspace{.7in}
%\includegraphics[scale=.5]{figure10m.ps}
%\hspace{.3in}
%\includegraphics[scale=.5]{figure10n.ps}\\
%\includegraphics[scale=.5]{figure10e.ps}
%\hspace{.3in}
%\includegraphics[scale=.5]{figure10f.ps}
%\hspace{.7in}
%\includegraphics[scale=.5]{figure10o.ps}
%\hspace{.3in}
%\includegraphics[scale=.5]{figure10p.ps}\\
%\includegraphics[scale=.5]{figure10g.ps}
%%%%%%%
\includegraphics[scale=.5]{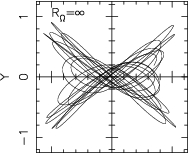}
\hspace{.3in}
\includegraphics[scale=.5]{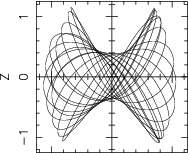}
\hspace{.7in}
\includegraphics[scale=.5]{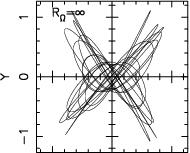}
\hspace{.3in}
\includegraphics[scale=.5]{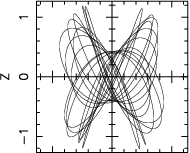}\\ 
\includegraphics[scale=.5]{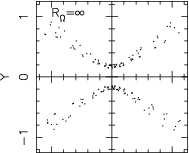}
\hspace{.3in}
\includegraphics[scale=.5]{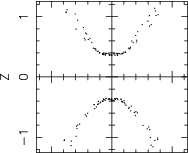}
\hspace{.7in}
\includegraphics[scale=.5]{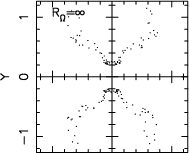}
\hspace{.3in}
\includegraphics[scale=.5]{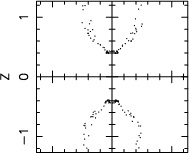}\\
\includegraphics[scale=.5]{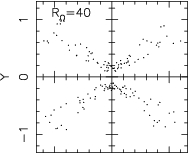}
\hspace{.3in}
\includegraphics[scale=.5]{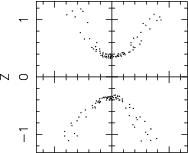}
\hspace{.7in}
\includegraphics[scale=.5]{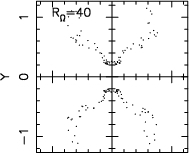}
\hspace{.3in}
\includegraphics[scale=.5]{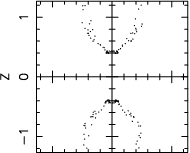}\\

\includegraphics[scale=.5]{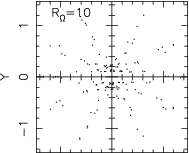}
\hspace{.3in}
\includegraphics[scale=.5]{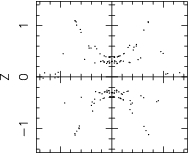}
\hspace{.7in}
\includegraphics[scale=.5]{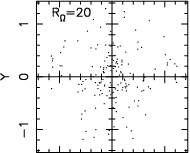}
\hspace{.3in}
\includegraphics[scale=.5]{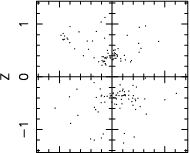}\\

\includegraphics[scale=.5]{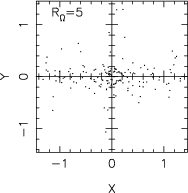}
\hspace{.3in}
\includegraphics[scale=.5]{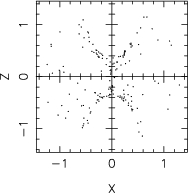}
\hspace{.7in}
\includegraphics[scale=.5]{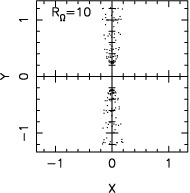}
\hspace{.3in}
\includegraphics[scale=.5]{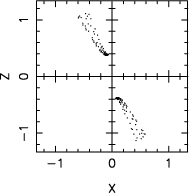}\\
\end{tabular}
\caption{Example of two inner long-axis ($x$) tube orbits. The top row
  shows two projections of the orbit in the $x$-$y$ (left) and $x$-$z$
  (right) planes in the stationary model. The second row shows
  cross sections of the same orbits with the $x$-$y$ and $x$-$z$
  planes. The next three rows show that the cross sections become
  broadened as the pattern frequency increases, becoming chaotic
  and/or tipped about the $y$-axis for higher pattern frequencies.
\label{fig:x-tubetip}}
\end{figure*}

Note that as the corotation radius decreases, inner long-axis ($x$)
tubes launched close to the center of the model (small $Y$ values)
become strongly chaotic, while those launched with intermediate $Y$
values are transformed into outer long-axis tubes and exhibit the
``anomalous'' behavior discussed by \citet{heisler_etal_82}
(i.e., they are tipped about the $y$-axis due to the Coriolis forces). The
destabilization of inner long-axis tubes is also a consequence of the
envelope doubling that arises due to the Coriolis
forces. Figure~\ref{fig:x-tubetip} shows projections as well as planar
cross sections of two inner long-axis tubes.  The two columns on the
left correspond to an orbit with a small pericenter radius which is
destabilized, while the orbit on the right goes through a phase where
it appears chaotic but is once again stable at a higher pattern
frequency, although it is tipped about the $y$-axis and significantly
flattened.

%%%%%%% Figure 9 %%%%%%%%%%
%%%%%%% Figure 9 %%%%%%%%%%

\citet{valluri_etal_10} showed that in triaxial $N$-body halos $x$-tubes
are the second most important family after the boxes ($\sim 85$\%) and
constitute 11\% of the orbits. They also showed that in prolate dark
matter halos \citep[the shapes most commonly found in cosmological
  simulations; ][]{dubinski_carlberg_91,jing_suto_00, allgood_etal_06}
the fraction of long-axis tubes can be as high as 75\%. Our finding that $x$-tubes with small pericenter radii are easily destabilized is
consistent with observations by \citet{valluri_etal_10} that long-axis
tubes with small pericenter radii are quite easily destabilized by the
presence of a central point mass in self-consistent (but non-rotating)
$N$-body models.
 
The discussions in this section indicate that as figure rotation
increases the fraction of phase space occupied by prograde $z$-tubes
and $x$-tubes (both clockwise and anti-clockwise) decreases and this
decrease is accompanied by an increase in the fraction of stable
retrograde $z$-tubes and chaotic orbits.

\subsection{Effects of Fast Rotation}
\label{sec:fast}

%%%%%%%% Figure 10 %%%%%%%%%
\begin{figure}
\centering
\begin{tabular}{c}
%\vspace{-1.cm}
\hspace{17pt}\includegraphics[scale=.4,angle=0]{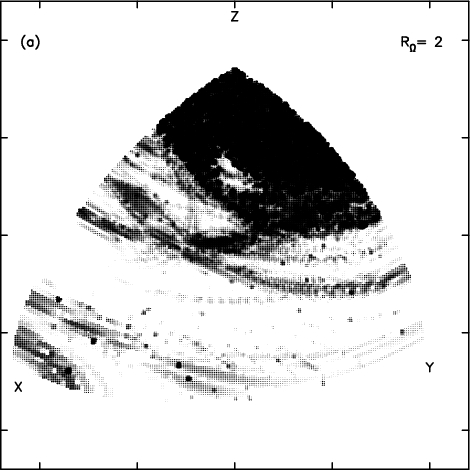}\\
\includegraphics[scale=.4,angle=0]{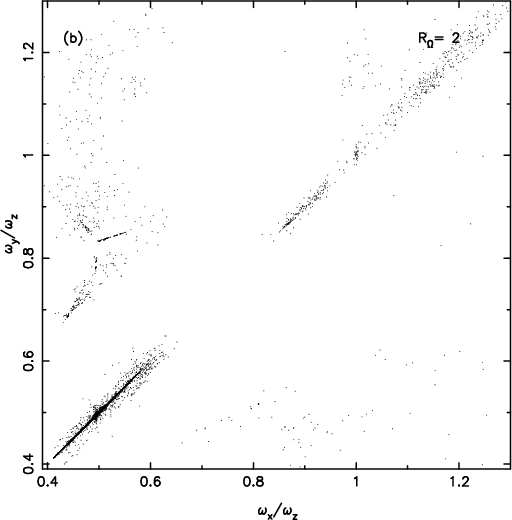}\\
\includegraphics[scale=.4,angle=0]{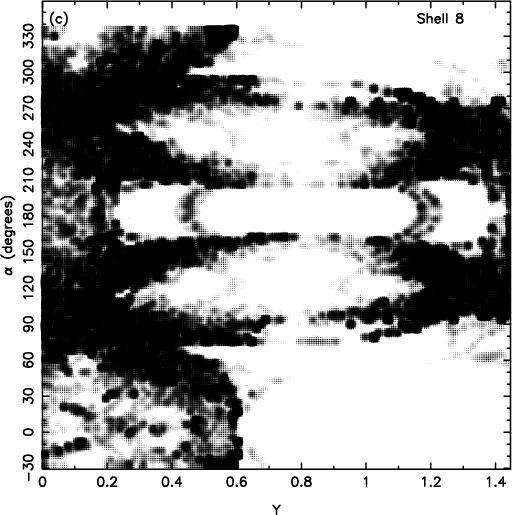}\\
%\hspace{17pt}\includegraphics[scale=.4,angle=0]{Box_diff_s8_r2.jpg}\\
%\includegraphics[scale=.4,angle=0]{Box_fmap_s8_r2.jpg}\\
%\includegraphics[scale=.4,angle=0]{Tube_s8_r2.jpg}\\
\end{tabular}
\caption{(a) Stationary start space diffusion map, (b)
  stationary start space frequency map, and (c) $Y$-$\alpha$ start
  space diffusion map for orbits in the 8th energy shell in models with
$c/a=0.5, T=0.58, \gamma = 1$, $M_{bh}= 0$, and $R_\Omega = 2$ ($R_\Omega/a_i=1.24$).
\label{fig:fast_maps}}
\end{figure}
%%%%%%%% Figure 10 %%%%%%%%%

We briefly explore the behavior of orbits when the pattern frequency
of figure rotation approaches that of bars. Most bars are rapidly
rotating with corotation radii roughly 1$-$1.4 times the length of
the bar \citep[e.g.,][]{debattista_etal_02,aguerri_etal_03,
  corsini_10,BT08}.

Figure~\ref{fig:fast_maps} shows the effect on orbits launched from
the 8th energy shell in the default model with figure rotation
corresponding to a corotation radius $R_\Omega= 2$
($R_\Omega/a_i=1.24$). Panel (a) shows the diffusion map for
orbits launched from the stationary start space in the default model;
in contrast with Figure~\ref{fig:box_diff_nobh}(d) ($R_\Omega=
5$), which was occupied almost entirely by chaotic boxlike orbits, at
a high pattern frequency the diffusion map shows a transition to
predominantly regular orbits. An inspection of the corresponding
frequency map (panel (b)) reveals that the orbits no longer have
box-like characteristics but instead predominantly lie along a
diagonal line indicating that they are more like $z$-tube orbits.
Panel (c) of Figure~\ref{fig:fast_maps} shows that while a
significant fraction of orbits are chaotic, the major tube families
(the $x$-tubes and $z$-tubes) do persist, and stable retrograde
$z$-tubes are again a dominant population.

\begin{figure}
\centering
\begin{tabular}{cc}
%\includegraphics[scale=.5]{boxs103_65.ps}\\
% from boxx104: orbits#10, 15, 28, 36,  56, 68,127,157,165, 170,    |180,190
\includegraphics[scale=.5]{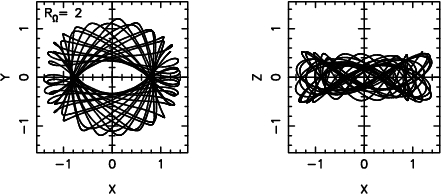}\\
\includegraphics[scale=.5]{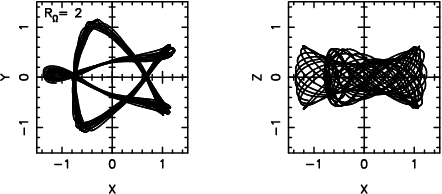}\\
\includegraphics[scale=.5]{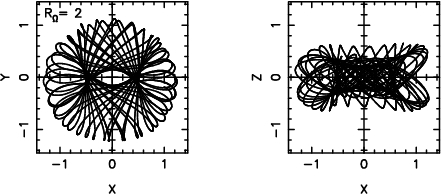}\\
\end{tabular}
\caption{Box orbits in models with high pattern speeds are transformed
  to regular retrograde orbits that circulate about the $z$-axis and
  loop around the center. Top row: a retrograde short-axis tube orbit
  that originates close to the $y$-axis on the equatorial plane that
  first appears at moderate pattern speeds ($R_\Omega= 10, 5$). Rows
  2$-$4: examples of regular ``bar-like'' loop-orbits that dominate the
  regular orbit population of orbits launched from the "box" start space
  at $R_\Omega=2$ ($R_\Omega/a_2=1.24$).}
\label{fig:yaxis_ztube} 
\end{figure}

Examples of some of the regular orbits launched from the box start
space that appear at the highest pattern frequency are shown in the
lower three rows of Figure~\ref{fig:yaxis_ztube}.  While these orbits do
loop around the $z$-axis and therefore superficially resemble
$z$-tubes they are in fact a distinct family more closely resembling
orbits in barred galaxies. This is not surprising since this pattern
speed, $R_\Omega/a_i = 1.24$, is in the range of values seen in fast
bars. 

As mentioned previously, the normalized angular momentum
$\langle{J_z/|J_z|}\rangle$ is close to unity for bona fide
$z$-tubes. However, these new loop-like orbits have a wide range of
$\langle{J_z/|J_z|}\rangle$ values, most of which are significantly
smaller than unity. This is because although these orbits loop about
the center on average in a retrograde sense, these orbits frequently
make small prograde ``epicycle-like'' loops (instead of turning points
in the stationary frame) (see left-hand plots of
Figure~\ref{fig:yaxis_ztube}. For these orbits
$\langle{J_z/|J_z|}\rangle$ is larger than the other two components of
angular momentum, but it is smaller than unity. Our automatic orbit
classifier therefore identifies these orbits as boxes and not
$z$-tubes. These orbits are not axisymmetric in the $x$-$y$ plane and
have elongated box-like projections in the $x$-$z$ plane which would be
useful in the construction of self-consistent triaxial models. 

%%%%% Figure 11 %%%%%%%%
%%%%% Figure 11 %%%%%%%%%%%%%%%

\newpage
\subsection{Dependence on radial shell}
\label{sec:radial}

In order to draw conclusions about the orbit composition and stability
of equilibrium triaxial galaxies with varying pattern speeds it is not
adequate to restrict the study to orbits with a single energy
(e.g., the 8th energy shell). It is important to assess how the
fractions of various orbit families depend on their potential energy,
in {\it self-consistent} models such as those constructed via
Schwarzschild's orbit superposition method. Self-consistent
Schwarzschild models will be constructed in a future study
\citep{valluri_11}. In the next two subsections, we will study
the properties of orbits at six different radii in a stationary model as
well as in a rotating model with $R_\Omega=10$.  Such an assessment,
while not a substitute for constructing self-consistent models, can
lead to insights into properties of such models. We consider orbits in
the default model for the "$Y$-$\alpha$" start space as well as orbits
from the "stationary" start space. For the latter we consider only a
model with a central point mass (representing an SMBH). Orbits launched
from start spaces equipotential surfaces corresponding to shells 2, 5,
8, 10, 12, and 16 (recall that outer edge of the 16th shell encloses 80\%
of the total mass of the model).  These shells have major axes (in
units of $a$) of 0.45, 0.95, 1.61, 2.22, 3.10, and 6.87, respectively. For
$R_\Omega= 10$, the corresponding values of $R_\Omega/a_i$ are $22.22,
10.52, 6.21, 4.50, 3.22,$ and $1.46$. For the 16th shell, the value of
$R_\Omega/a_i = 1.46$ is approaching the value of that seen in fast
bars, accounting for the increased fraction of the new family of
regular loop-like orbits.

\subsubsection{Tubelike orbits}
\label{sec:radial_tube}

\begin{figure*}
\centering
\begin{tabular}{cc}
\includegraphics[scale=.3,angle=0]{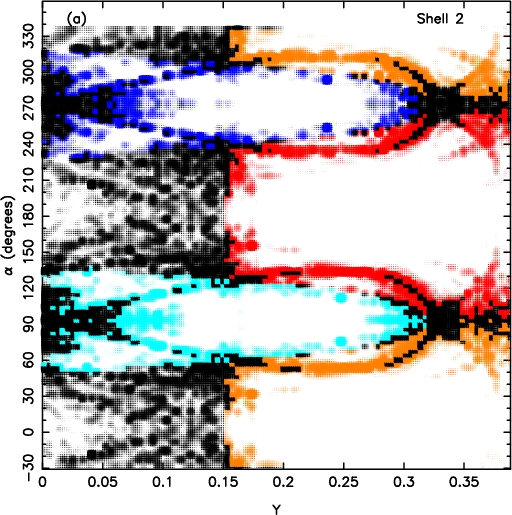}
\includegraphics[scale=.3,angle=0]{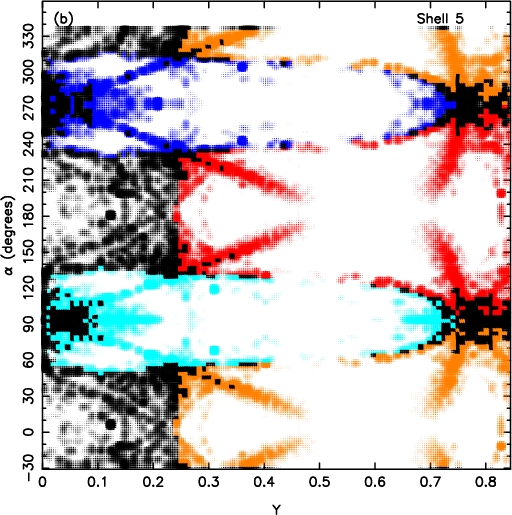}\\
\includegraphics[scale=.3,angle=0]{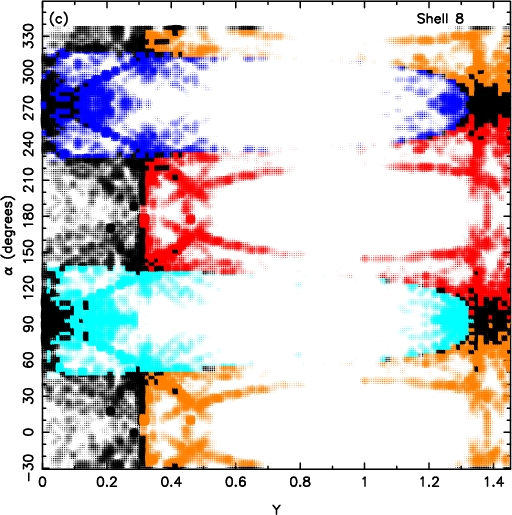}
\includegraphics[scale=.3,angle=0]{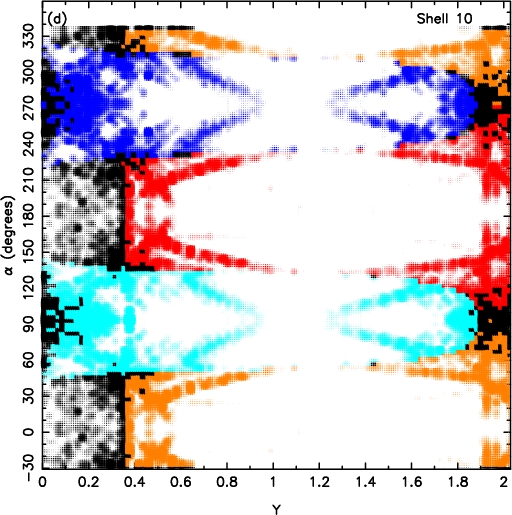}\\
\includegraphics[scale=.3,angle=0]{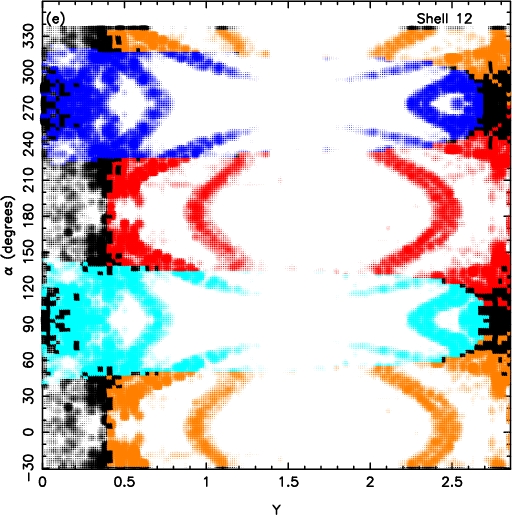}
\includegraphics[scale=.3,angle=0]{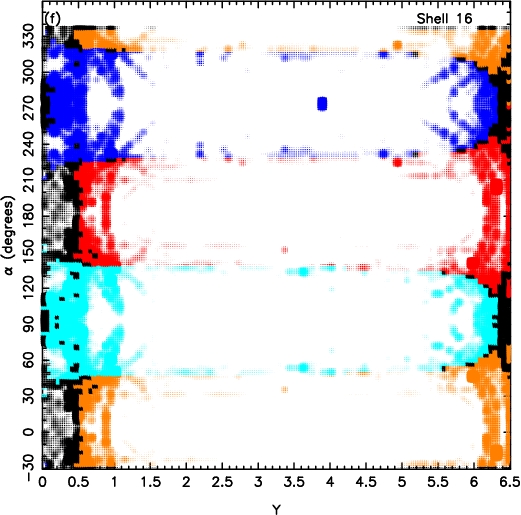}\\
\end{tabular}
\caption{Diffusion maps on the $Y$-$\alpha$ start space for orbits
  at six different energy levels in a non-rotating model with ($c/a=0.5,
  T= 0.58, \gamma = 1,$ and $M_{bh}=0$).  
\label{fig:norot_radius}}
\end{figure*}

Figure~\ref{fig:norot_radius} shows diffusion maps of the $Y$-$\alpha$
start space for the six different shells. The maps are color coded as in
Figure~\ref{fig:yalpha_fmap_diff_gamma_0}(b) with black denoting
box-like orbits, cyan, and blue denoting clockwise and anti-clockwise
$x$-tubes, ochre denoting clockwise (retrograde) $z$-tubes and red
denoting anti-clockwise (prograde) $z$-tubes.

The first five shells show many of the characteristics that we saw
previously: box orbits and chaotic orbits dominate at small $Y$ values
in all shells. In shell 2, box orbits and inner long-axis tubes are
found within the inner half of the diffusion plot and are often mildly
chaotic. In the outer parts of the model (shell 16), the box orbits
only occupy a small fraction of the phase space and the tube orbit
families (clockwise and anti-clockwise varieties) occupy equal areas
of the initial condition space. As before the various orbit families
are separated by chaotic separatrix layers and in addition unstable
resonances appear as dark bands in the regions occupied by $z$- and
$x$-tubes.

%%%%% Figure 12 %%%%%%%%%%%%%%%%%%%%%%%
%%%%%%%%%% Figure 12 %%%%%%%%%%%%%%%%%

\begin{figure*}
\centering
\begin{tabular}{cc}
\includegraphics[scale=.3,angle=0]{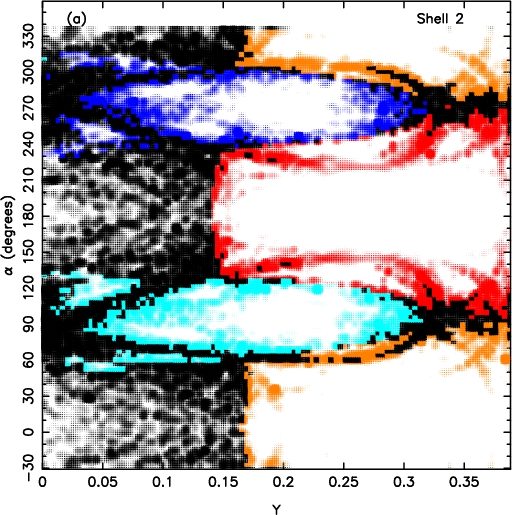}
\includegraphics[scale=.3,angle=0]{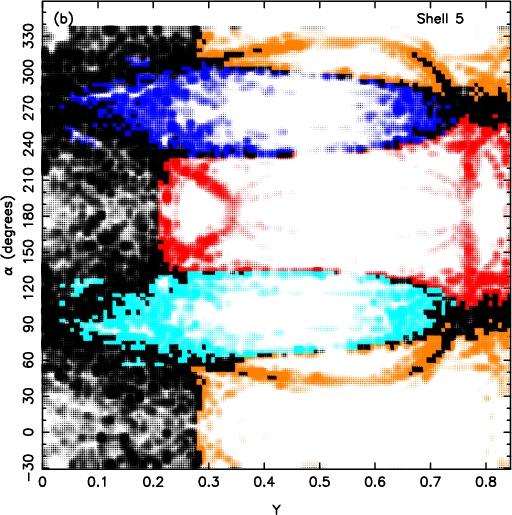}\\
\includegraphics[scale=.3,angle=0]{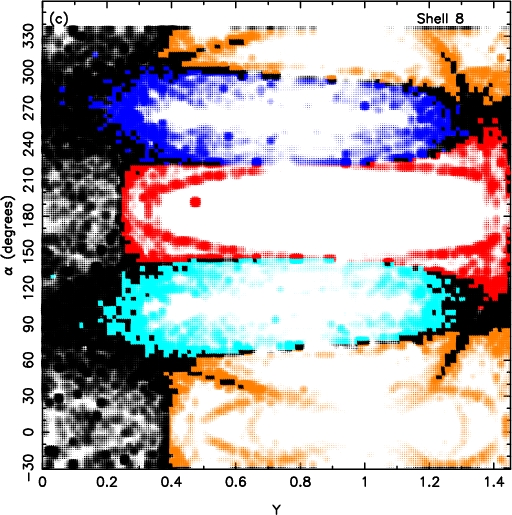}
\includegraphics[scale=.3,angle=0]{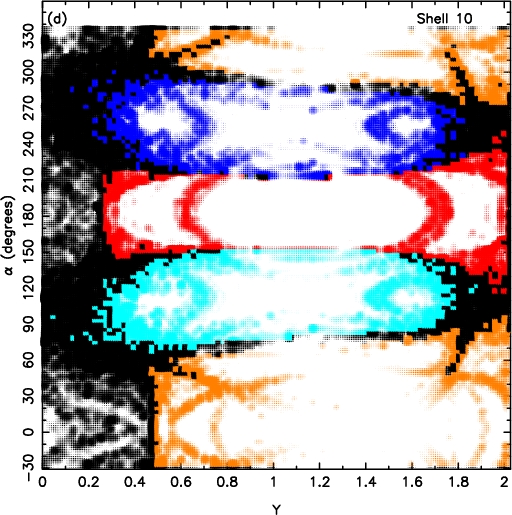}\\
\includegraphics[scale=.3,angle=0]{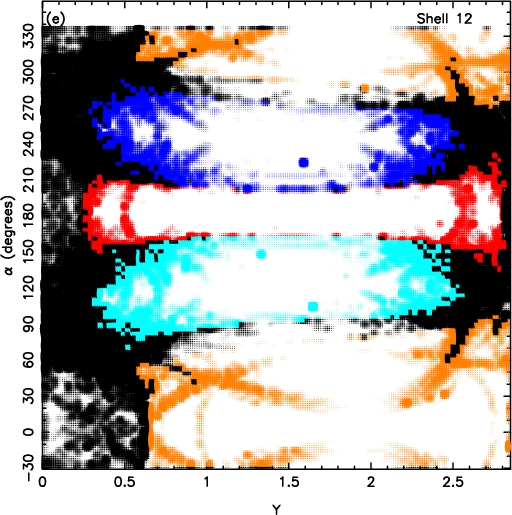}
\includegraphics[scale=.3,angle=0]{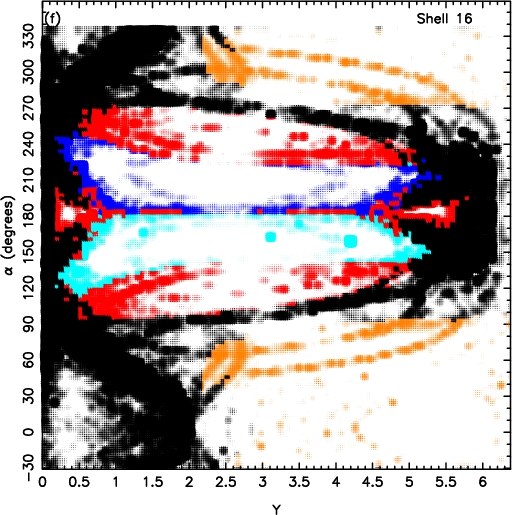}\\
\end{tabular}
\caption{Diffusion maps on the $Y$-$\alpha$ start space for orbits
  at six different energy levels in a model with ($c/a=0.5, T= 0.58,
  \gamma = 1,$ and $M_{bh}=0$). For $R_\Omega=10$, $R_\Omega/a_i:
  22.22, 6.21,5.98, 3.22,$ and $1.54$ for radial shells 2, 5, 8, 10, 12, and 16,
  respectively.
\label{fig:withrot_radius}}
\end{figure*}

Figure~\ref{fig:withrot_radius} shows the effect of figure rotation on
the orbits in Figure~\ref{fig:norot_radius}. Note that the corotation
radius of $R_\Omega=10$ was selected because this corotation radius
lies outside shell 16 ensuring that none of the shells experience the
fast pattern speeds seen in Section~\ref{sec:fast}. Only $\sim$10\% of the
total mass of the model lies outside a corotation radius of
$R_\Omega=10$.

%%%%%%%%%% Figure 13 %%%%%%%%%%%%%%%%%
%%%%%%%%%% Figure 13 %%%%%%%%%%%%%%%%%

Figure rotation produces roughly the same qualitative effects on
orbits in all six radial shells although the details depend on radius:
(1) the fraction of chaotic (and boxlike) orbits (black/intense
color) increases relative to the non-rotating model, especially in the
outer shells, (2) the fraction of $x$-axis tubes (blue/cyan)
decreases, (3) in shell 8 and beyond the region occupied by
prograde $z$-tubes (red) shrinks (there are no equatorial plane
prograde $z$-tubes by shell 16 but two new bands appear around
$\alpha\sim 90^\circ, 270^\circ$)\footnote{The new bands of prograde
  $z$-tubes that appear at $\alpha \sim 100$ and $\alpha \sim 260$ at
  the outer most energy shells are classified by our automatic
  classifier as $z$-tubes but are in fact $x$-tubes that have been
  tipped about the $y$-axis by nearly $90^\circ$ as predicted by
  \citep{heisler_etal_82}.}, and (4) the retrograde $z$-tubes (white
regions surrounded by ochre bands) dominate.

\begin{figure*}
\centering
\begin{tabular}{cc}
\includegraphics[scale=.4,angle=90]{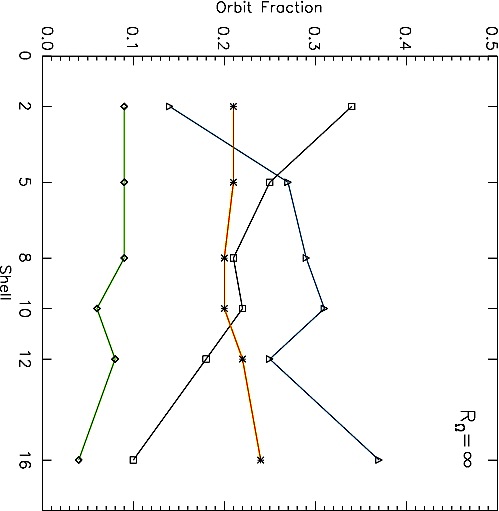}
\includegraphics[scale=.4,angle=90]{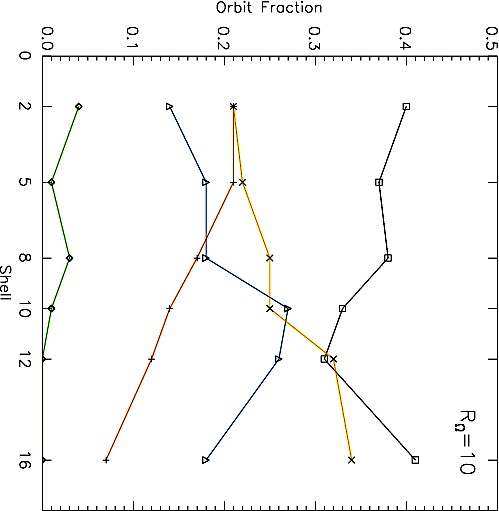}\\
\end{tabular}
\caption{Fractions of orbits in major families as a function of radial
  energy shell number in the default models, left: without
  figure rotation and right: with figure rotation.  Squares
  ($\Box$) connected by black lines denote box-like orbits (which are
  predominantly chaotic boxes); triangles ($\triangle$) connected by
  blue lines denote outer $x$-axis tube-like orbits; diamonds
  ($\diamond$) connected by green lines denote inner $x$-axis tubes;
  crosses ($\times$) connected by red lines denote prograde $z$-tubes,
  and plus signs ($+$) connected by ochre lines denote retrograde
  $z$-tubes. In the absence of figure rotation (left) the
  fractions of prograde and retrograde $z$-tubes are exactly the same
  so the ($\times$) and ($+$) symbols overlap; at a higher pattern speed
  (right), the ochre curve (retrograde $z$-tubes) and red curve
  (prograde $z$-tubes) separate clearly showing the dominance of the
  retrograde family, especially at large radii.
\label{fig:orbit_fractions}}
\end{figure*}

To better quantify the effects of figure rotation on the orbit
populations we computed fractions of orbits in each of the major
families. As discussed previously, we use time-averaged normalized
angular momentum values to distinguish between boxes, $x$-tubes, and
$z$-tubes. In addition, we visually inspected 225 uniformly distributed
orbits in each shell to confirm these automatic classifications and to
also allow us to distinguish between inner and outer $x$-tubes. The
results of the orbit classification are plotted in
Figure~\ref{fig:orbit_fractions} for the models in
Figure~\ref{fig:norot_radius} ($R_\Omega=\infty$) and
Figure~\ref{fig:withrot_radius} ($R_\Omega=10$). In the absence of
figure rotation (left) the fractions of prograde and retrograde
$z$-tubes are exactly the same so the ($\times$) and ($+$) symbols
overlap; at a higher pattern speed (right), the ochre curve
(retrograde $z$-tubes) and red curve (prograde $z$-tubes) separate
clearly showing the dominance of the retrograde family, especially at
large radii.  In this plot we do not distinguish between chaotic
and boxlike orbits, since to do so would require setting an arbitrary
``cutoff'' value for the diffusion parameter at which orbits would be
classified as chaotic. Most of the orbits classified as boxlike are in
fact chaotic, but a large number of the tube like orbits (especially
the $x$-axis tubes) are quite strongly chaotic as can be seen from the
intensity of the blue and cyan regions in
Figures~\ref{fig:norot_radius} and~\ref{fig:withrot_radius}.

%%%%%%%%%% Figure 14 %%%%%%%%%%%%%%%%
%%%%%%%%%%% Figure 14 %%%%%%%%%%%%%

The figure shows that the fraction of box-like (chaotic) orbits
increases slightly in the inner shells but increases significantly in
the outer shells (but is as high as 40\% in many shells in the
rotating models). Note that in contrast, the fraction of box orbits
drops with radius in the stationary model (left). The fraction of
inner $x$-axis tubes is not very significant even in the stationary
model but decreases significantly, disappearing entirely at larger
radii in the rotating model. (While inner long-axis tubes do not play
a significant role in maximally triaxial models, they are elongated
along the major axis of the potential and are vital, in addition to
box orbits, for maintaining the triaxial shape in significantly
prolate models.)  The outer long ($x$) axis tube fraction decreases
slightly at most energies, but especially at large radii. The fraction
of prograde versus retrograde $z$-tubes remains almost unchanged in
the inner two shells, but there is a significant decrease in fraction
of prograde $z$-tubes (red curve (+) signs) in the outer energy shells
and a corresponding increase in the retrograde fraction (ochre curve
$\times$). We note that the behavior of orbits in shell 8 in response to
figure rotation is broadly indicative of average behavior of orbits
in different radial shells. Thus, it is reasonable to use this shell to
gauge the average effect of figure rotation on the entire model, but
not to predict the detailed radial structure of such galaxies.

\subsubsection{Box-like orbits}
\label{sec:radial_box}

Since box orbits are generally regarded as the most important family
for supporting triaxiality, it is also necessary to study the radial
dependence of their stability. The stability of box orbits is strongly
affected by chaotic scattering by a central black hole. Since
SMBHs are now believed to reside at the centers of all galaxies with a
significant bulge/ellipsoidal components, we now consider behavior of
orbits launched from the stationary (box) start space in models with SMBHs.

%%%%%%%%%%% Figure 15 %%%%%%%%%%%%%%
\begin{figure*}
\centering
\begin{tabular}{cc}
\includegraphics[scale=.3,angle=0]{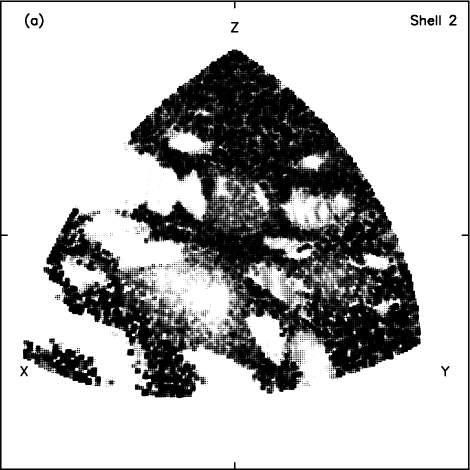}
\includegraphics[scale=.3,angle=0]{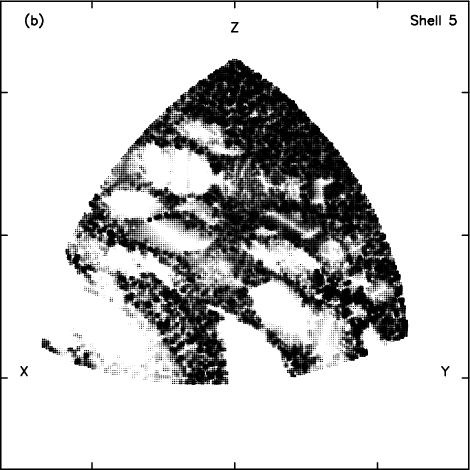}\\
\includegraphics[scale=.3,angle=0]{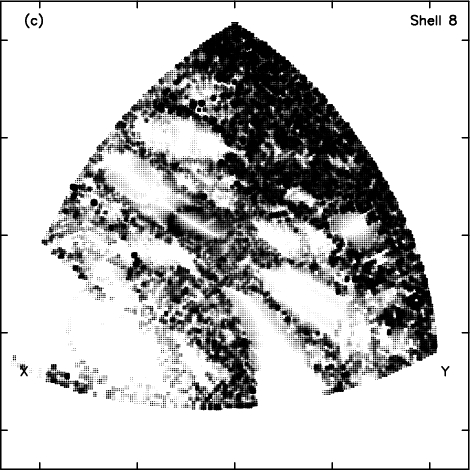}
\includegraphics[scale=.3,angle=0]{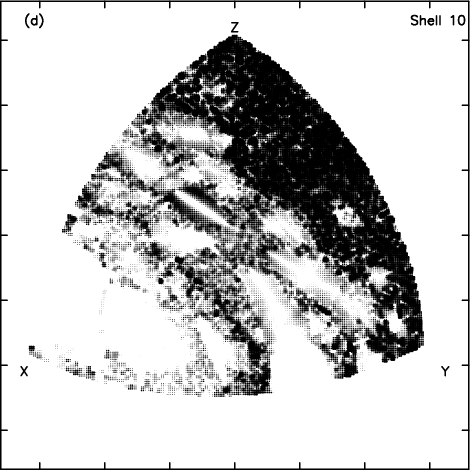}\\
\includegraphics[scale=.3,angle=0]{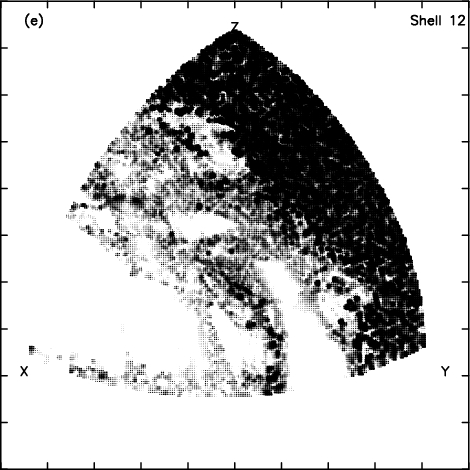}
\includegraphics[scale=.3,angle=0]{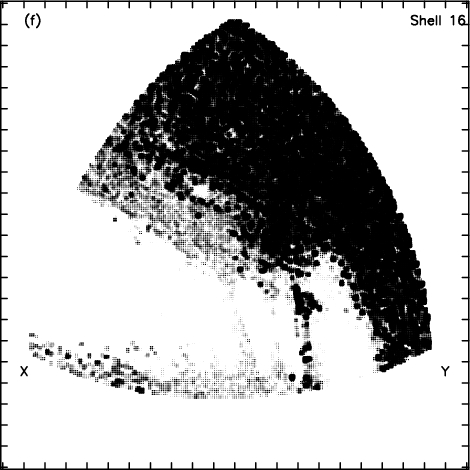}\\
\end{tabular}
\caption{Diffusion maps on the box start space for orbits at six
  different radial energy shells (as labeled) in a model with
  ($c/a=0.5, T= 0.58,$ and $\gamma = 1$). The model has a central black hole
  of $M_{bh}=0.001$) and no figure rotation.
  \vspace{.3cm}
\label{fig:norot_box_radius}}
\end{figure*}
%%%%%%%%%%%% Figure 15 %%%%%%%%%%%%%

Figure~\ref{fig:norot_box_radius} shows diffusion maps for orbits launched from six different radial energy shells (as labeled) in the absence of figure rotation.  There is a large fraction of chaotic orbits at all radii. At small and intermediate radii, there are numerous small resonant islands. In the outer two shells, there are only three larger islands associated with regular orbits and these are mostly associated with non-resonant box orbits (although some small resonant islands persist).

%%%%%%%%%%% Figure 16 %%%%%%%%%%%%%%
\begin{figure*}
\centering
\begin{tabular}{cc}
\includegraphics[scale=.3,angle=0]{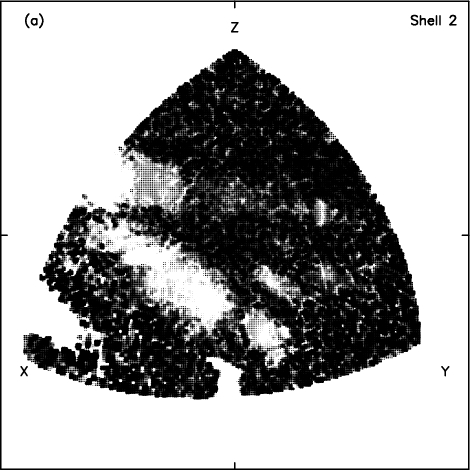}
\includegraphics[scale=.3,angle=0]{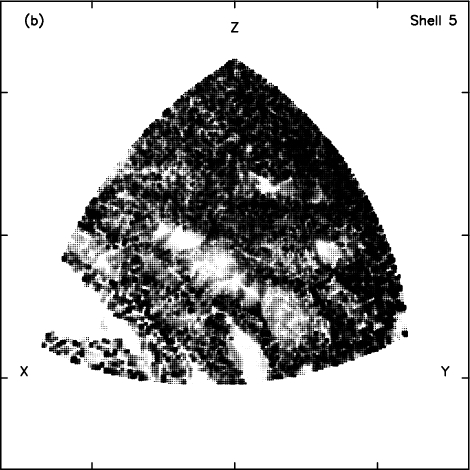}\\
\includegraphics[scale=.3,angle=0]{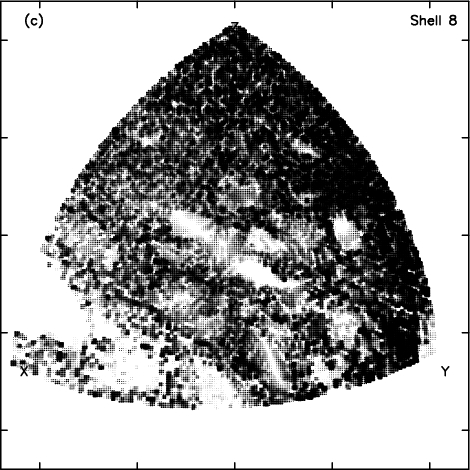}
\includegraphics[scale=.3,angle=0]{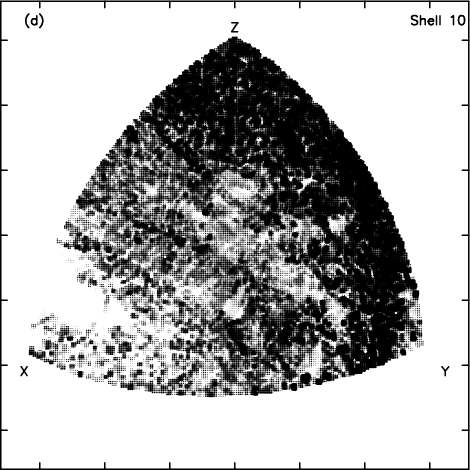}\\
\includegraphics[scale=.3,angle=0]{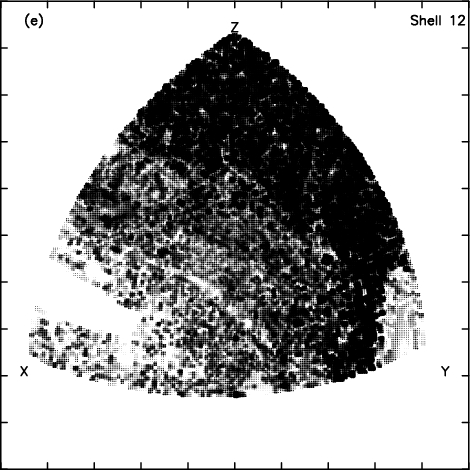}
\includegraphics[scale=.3,angle=0]{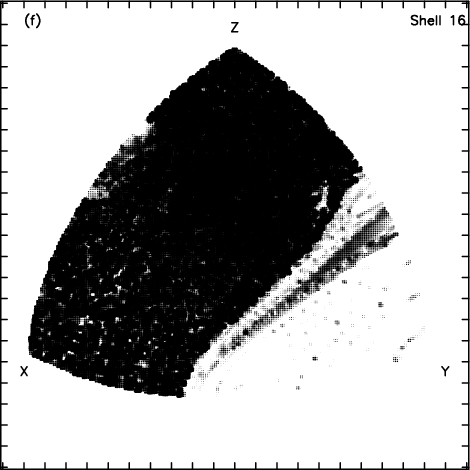}\\
\end{tabular}
\caption{Diffusion maps on the box start space for orbits
  at six different energy levels in a model with ($c/a=0.5, T= 0.58,
  \gamma = 1,$ and $M_{bh}=0.001$) with co-rotation radius of $R_\Omega = 10$. For shells 2,5, 8,10,12, and 16, respectively, this pattern speed corresponds to $R_\Omega/a_i: 22.22, 10.52, 6.21,4.5, 3.22,$ and $1.46.$
\label{fig:box_withrot_radius}}
\vspace{.3cm}
\end{figure*}
%%%%%%%%%%%% Figure 16 %%%%%%%%%%%%%

Figure~\ref{fig:box_withrot_radius} shows the effect of figure rotation ($R_\Omega = 10$) on the box-like orbits in Figure~\ref{fig:norot_box_radius}.
In the inner three shells we see that figure rotation has the effect of reducing the sizes of the regular islands of resonant orbits that we previously discussed in Section~\ref{sec:box}. The disappearance of resonant islands is particularly notable in the inner four shells where the non-rotating model (Figure~\ref{fig:norot_box_radius}) shows numerous small islands. In the outer two shells (which are closer to the corotation radius), we see the onset of regularity associated with orbits close to the $y$-axis. These orbits have $z$-tube-like characteristics which was previously demonstrated  in Figures~\ref{fig:yaxis_ztube} and~\ref{fig:fast_maps}. 

%%%%%%%%%%%START OF CHANGES%%%%%%%%%%%%%%

Very near the SMBH, within its gravitational sphere of influence,
motion in non-rotating triaxial galaxies is essentially regular;
orbits take the form of perturbed Keplerian ellipses that 
gradually precess due to the torques from the stellar potential
\citep{merritt_vasiliev_10}.
The innermost radial shell of our models contains a stellar
mass $\sim 0.05M_\mathrm{gal}\approx 50M_\mathrm{bh}$ and
so lies well outside of the influence sphere.
%%%%%%%%%%%END OF CHANGES%%%%%%%%%%%%%%%

\subsection{The effect of varying the shape of the model}
\label{sec:shape}

The projected shape of a galaxy on the sky depends both on its
intrinsic shape (i.e., the value of $T$ and $c/a$) as well as its
orientation relative to the line of sight. Although it is possible to
measure only the projected shape of an individual early-type galaxy it
is possible to infer the properties of the distribution of shapes from
the assumption that each individual in a large sample should be
randomly oriented. Using surface photometry of the apparent shapes of
elliptical galaxies it has been shown that their intrinsic shapes
cannot be explained if they are exclusively a population of randomly
oriented oblate spheroids, but the distribution of apparent shapes is
consistent with their being a population of triaxial ellipsoids
\citep{fasano_vio_91,
  lambas_etal_92,ryden_92,tremblay_merritt_95,vincent_ryden_05}. Observationally,
it appears that the lower luminosity ellipticals tend to be flatter
and more consistent with disky oblate systems while the higher
luminosity systems are more likely moderately
oblate triaxial. Strongly prolate, perfectly oblate systems, or
perfectly spherical systems are rare and the more luminous elliptical
galaxies are probably oblate triaxial. The models we investigated in
previous sections are close to maximally triaxial and consequently
could be somewhat unrealistic. \citet{merritt_97} explored the range
of self-consistent (non-rotating) triaxial galaxies of various shapes
that could be constructed exclusively with regular orbits, in Dehnen
models with $\gamma=2$. He showed that the allowed range of shapes for
which all orbits were regular tended to be mostly oblate and
oblate triaxial with a small number of nearly prolate shapes allowed.

\begin{figure*}
\centering
\begin{tabular}{cc}
\includegraphics[scale=.4,angle=0]{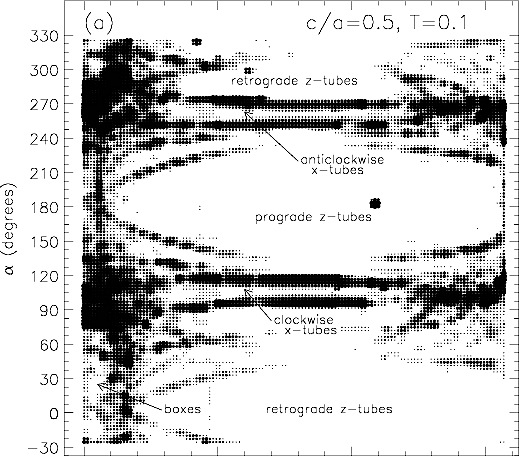}
\includegraphics[scale=.4,angle=0]{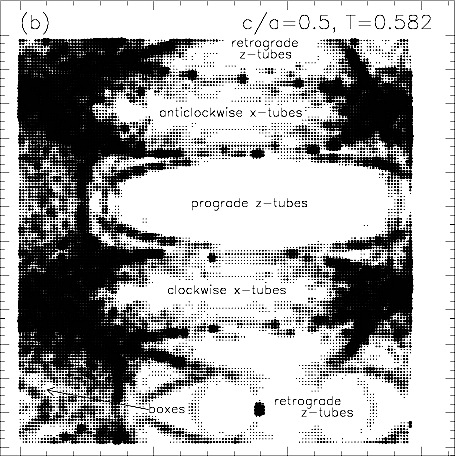}\\
\includegraphics[scale=.4,angle=0]{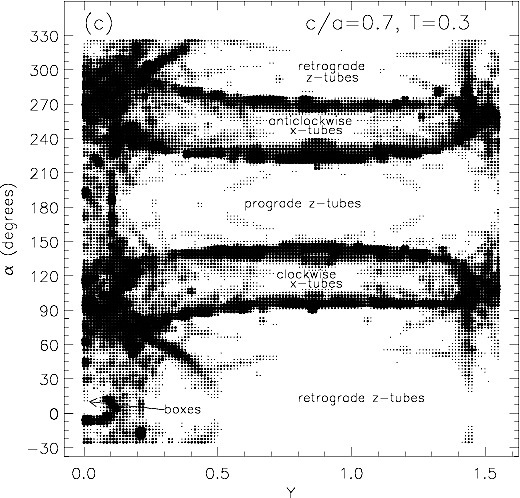}
\includegraphics[scale=.4,angle=0]{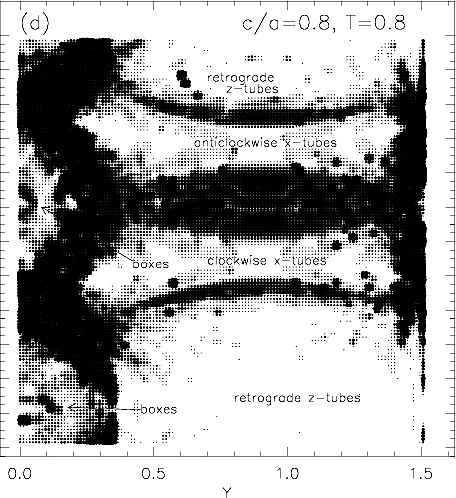}\\
\end{tabular}
\caption{Effect of varying the shape of the galaxy on the orbit
  population is seen in $Y$-$\alpha$ diffusion maps for models with
  ($\gamma = 1$ and $R_\Omega = 10$).  Panels correspond to models
  with various shapes as indicated by values of $c/a$ and $T$. The
  model (a) is nearly oblate, (b) is nearly maximally
  triaxial, (c) is moderately triaxial, and (d) is nearly
  prolate. The different orbit families are clearly separated by
  stochastic separatrix layers.
\label{fig:shape}}
\end{figure*}

It is out of the scope of this paper to do a full exploration of the
dependence of orbital structure on galaxy shape. However, we studied
the effect of figure rotation ($R_\Omega=10$) on orbits launched from
shell 8 in models with three additional shapes, with $\gamma=1$ and no black
hole.  Figure~\ref{fig:shape} shows the $Y$-$\alpha$ diffusion maps for
four different models: (1) a nearly oblate model, (2) the
nearly maximally triaxial model studied in previous sections, (3) a moderately triaxial model, and (4) a nearly prolate
model.

In the nearly oblate model (panel (a)) the fraction of $x$-axis
tubes is insignificantly small, being almost entirely replaced by
$z$-tubes. The few surviving $x$-tubes are outer long-axis tubes. Box
orbits continue to occupy only a small fraction of this start
space. This is the only rapidly rotating model in which there is not a
large difference between the prograde and retrograde short-axis tube
orbit fractions (retrograde fraction=0.42, prograde fraction=0.35).
The maximally triaxial model in panel (b) has a larger fraction
of chaotic orbits than models of any other shape. The moderately
triaxial model (panel (c)) has a significant fraction of $z$-axis
tubes (especially retrograde ones) and a moderate fraction of
long-axis ($x$) tubes. Almost all the long-axis tubes are associated
with anomalous orbits and the few that show characteristics of inner
long-axis tubes are mildly chaotic. Most of the box orbits in the
model are quite stable.  Finally, the nearly prolate model (panel (d)) is dominated by retrograde $z$-axis tubes while the region
occupied by prograde $z$-tubes is now entirely occupied by chaotic
orbits. Box orbits and $x$-tubes are found in smaller numbers and some
are moderately chaotic.

Admittedly, it is dangerous to draw conclusions about the shape
distributions of self-consistent galaxy models with figure rotation,
from looking at the populations of orbits in a selected number of
start spaces. However, our analysis shows that orbit populations
evolve slowly and systematically within an individual model (with
varying $\alpha$, $Y$, and radial shell), as well as with the shape of
the figure and the degree of figure rotation. In particular we see
that in models with a range of shapes, stable short-axis tubes with
retrograde motion are the dominant population in the start space. At
fast pattern speeds, box orbits are also replaced by retrograde loop-like orbits that circulate about the $z$-axis. The dominance of
retrograde $z$-tubes at moderate to fast pattern speeds suggests that
these orbits are very likely to be important in self-consistent
triaxial galaxies.

%%%%%%%%%%%% Figure 17 %%%%%%%%%%%%%%
%%%%%%%%%%% Figure 17 %%%%%%%%%%%%%%%%

\section{Summary and Discussion}
\label{sec:discuss}

%%%%%%%%%%% MINOR CHANGES THROUGHOUT THIS SECTION %%%%%%%%%%
%%%%%%%%%%% (ORIGINAL TEXT COMMENTED OUT) %%%%%%%%%%%%%%%

We have investigated the effects of figure rotation on the orbital
structure of a family of triaxial galaxy models 
%\citep{dehnen_93} (THEY AREN'T REALLY DEHNEN'S MODELS) 
that provide an accurate representation of the luminosity 
%density (REDUNDANT)
profiles of elliptical galaxies. 
%Our goal was to explore the behavior of orbits
%in a family of triaxial galaxies when subjected of figure rotation
%about the short axis of the models. (MOSTLY REDUNDANT)
Most of our study is restricted to
models that are close to maximally triaxial with $c/a=0.5, T= 0.58$,
with shallow cusps of $\gamma=1$.
Rotation of the figure was assumed to be about the short axis.

Our major results are summarized below.
 
\begin{enumerate}

%\item Orbits in the slowly rotating triaxial models  ($\Omega_p \gtrsim 1.44$\kmskpc) are %similar to those in stationary elliptical galaxies when
%  examined in the frame co-rotating with the pattern speed of the
%  figure. Figure rotation produces both centrifugal forces and
%  coriolis forces whose effect is more significant at higher pattern
%  speeds. (I DON'T THINK WE NEED TO SAY THESE THINGS -- 
%THEY ARE PRETTY OBVIOUS.)

\item As figure rotation increases, the fraction of initial condition
space occupied by prograde $z$- and $x$-tube orbits (both clockwise and 
anti-clockwise) decreases.
The fraction of stable retrograde $z$-tubes and chaotic orbits
correspondingly increases.

%  \item The addition of a central point mass representing a SMBH
%  increases the degree of destabilization of the box and
%  inner long-axis tube families, but the qualitative dependence on
%  pattern speed is similar to that found in models with a central cusp
%  slope of $\gamma=1$ and no SMBH. Increasing the central cusp slope
%  to $\gamma=2$ has a similar effect to the addition of a central
%  point mass. (DITTO)

\item Almost all regular 
%box like 
 box-like orbits in realistic triaxial
  potentials with central cusps and/or black holes are associated with
  resonant orbits.  The 
%coriolis 
  Coriolis forces in the rotating frame of a
  triaxial potential 
%destabilizes 
 destabilize these near-resonant box orbits as well as
  inner long-axis tube orbits, by thickening them enough to drive them
  into the 
%destabilizing 
  center. The outer long-axis tubes are also
  slightly destabilized by figure rotation but less than the inner
  long-axis tubes. Box orbits are further destabilized because the
  modulation of the orbital frequencies by the rotational frequency
  results in a shrinking in the overall range of orbital frequencies,
  and therefore in significant resonance overlap.
% - a well known cause of global chaos \citep{chirikov_79}. 
  Orbits that are chaotic in the
  stationary model are not in general stabilized unless they occupy
  very special regions in phase space.

\item Over a wide range of pattern speeds there is little evidence
  that orbits that are chaotic in the stationary model are stabilized
  in the rotating model 
%by deflection about the center. 
  due to avoidance of the center.
  At fairly
  rapid rotation speeds, a small region close to the $y$-axis that is
  occupied by chaotic box orbits in the stationary model is indeed
  stabilized by figure rotation as predicted by
  \citet{gerhard_binney_85}. However, this stabilization is due to a
  complete transformation of the orbital characteristics from box-like
  to tube-like, rather than due to deflection of the chaotic box orbit
  around the destabilizing center. At even higher pattern speeds
  (Figure~\ref{fig:fast_maps}) $R_\Omega/a_i \lesssim 3$ box-like orbits
  begin to show a sudden transformation to a new 
%families 
  family of
  retrograde 
%loop like 
  loop-like orbits 
%seen in rapidly rotating bars.  
  similar to those seen in rapidly rotating bars. These
  orbits loop around the $z$-axis in an overall retrograde sense, but
  execute small ``epicycle''-like prograde motions at points in the
  orbit that correspond to stationary points in the non-rotating
  model.

\item A surprising result of our study is that 
 as the pattern speed $\Omega_p$ increases there is a
decrease in the overall fraction of prograde short-axis tubes which are
increasingly replaced by the retrograde variety. Prograde short-axis tubes only persist when launched at angles $\sim \pm 20^\circ$ from the
equatorial plane of the model. Orbits launched at larger angles to the
equatorial plane have lower angular momenta about the short axis at a
given energy, and consequently tend to ``fall behind'' the figure and
become retrograde as it rotates past them.

\item There is a fairly strong dependence 
%on
  of 
  the changes in the
  behavior of orbits on the radial energy level from 
%they are
  which they are launched. 
  (1) The fraction of inner $x$-tubes decreases slightly at
  all radii, essentially disappearing at large radii. (2) The fraction
  of outer $x$-tubes also decreases at all radii. (3) The fraction of
  retrograde $z$-tubes increases and is accompanied by a decrease in
  the fraction of prograde $z$-tubes. (4) There is an overall increase
  in the fraction of chaotic/box-like orbits at all radii, but the
  increase is most significant at intermediate and large radii. 
%The
%  behavior of orbits in shell 8 represents the average behavior at all
%  radii. (NOT SURE WHY WE NEED THAT SENTENCE.)

\item As the shape of the triaxial figure changes, the orbital
  populations at 
%shell 8
  shell 8 (roughly the half-mass radius)
   vary. Yet in all cases with figure rotation
  (with $R_\Omega=10$) retrograde $z$-tubes are the most robust and
  stable population. The fraction of chaotic orbits is lower than in
  the maximally triaxial models.
\end{enumerate}

\bigskip
 
It has been known for three decades, since the work of
\citet{schwarzschild_82}, that self-consistent triaxial equilibria can
be constructed if the triaxial figure is rotating slowly.
Schwarzschild's study was restricted to galaxies with central cores;
our study of orbits in more realistic triaxial models supports the
view that galaxies with slow figure rotation have a large fraction of
stable ``triaxial orbit families'' such as boxes and long-axis tubes
which can be used to construct stable triaxial models. We find,
however, that as the pattern speed of figure rotation increases, these
two important families (stable resonant boxes and long-axis tubes)
become increasingly unstable, due to the effects of envelope
doubling. When the pattern frequency is between $\Omega_p \simeq
3$-$30$~\kmskpc (for an elliptical galaxy of scale radius 5~kpc and
mass $5\times 10^{11}M_{\odot}$), figure rotation produces a sufficiently
large amount of global chaos to make it unlikely that equilibrium
triaxial models with such pattern speeds would 
%exist
exist $-$ at least, if one assumes that a significant fraction of elongated regular orbits are required. 
At higher
pattern frequencies, comparable to those of fast bars, the phase space
associated with box-like orbits becomes stable once again, although the
orbits now resemble loops found in fast bars.

%The greatest uncertainty with constructing self-consistent triaxial
%galaxy models with figure rotation is with respect to the expected
%pattern speeds. 
Pattern speeds of real triaxial galaxies and dark-matter halos are 
poorly constrained.
We explored a range of values for figure rotation
motivated by theoretical measurements of pattern speeds of dark matter
halos arising from cosmological simulations
\citep{bailin_steinmetz_04,bryan_cress_07} as well as one claimed
observational measurement of rotation of a triaxial dark matter halo
with a pattern speed in NGC~2915 \citep{bureau_etal_99} and an early-type galaxy NGC~2794 \citep{jeong_etal_07}. The shape assumed for the
rigid triaxial dark matter halo potential in the model for NGC 2915
\citep{bekki_freeman_02} ($c/a= 0.6, T= 0.56$) is close to our default
model shape ($c= 0.5, T= 0.58$). Our study suggests that the pattern
speed of $\sim 7\pm 1$~\kmskpc required to produce the extended spiral
features in this galaxy is adequate to significantly destabilize the
two major families that sustain the high density along the major axis,
namely, the box orbits (predominantly associated with resonances) as
well as the inner long-axis tubes. This does not imply that the spiral
arms in this galaxy are not generated by tidal torques from a triaxial
halo but does suggest that the halo is unlikely to be as strongly
triaxial as assumed by these authors.  A rapidly rotating
prolate-triaxial halo \citep[predicted by cosmological $N$-body
  simulations,][]{dubinski_carlberg_91,jing_suto_00, allgood_etal_06})
is also unlikely, because inner long-axis tubes, which are important
in such models \citep{valluri_etal_10}, are easily destabilized by
large pattern speeds.

%It has been previously argued \citep{merritt_97} that while a wide
%range of theoretical shapes are possible for elliptical galaxies, only
%nearly oblate or nearly prolate models can be constructed with only
%regular orbits. It has been shown that triaxial galaxies with central
%supermassive black holes can have up to 50\% chaotic orbits and still
%remain stable for extended periods of time \citep{poon_merritt_04}.
If the orbits that are required to construct triaxial galaxies, namely,
the boxes and inner long-axis tubes, become highly chaotic (or
disappear entirely) at intermediate pattern speeds, they are less
elongated than necessary to construct self-consistent galaxies
\citep{merritt_valluri_96}. This places additional constraints on both
the range of shapes as well as the range of pattern speeds of real
elliptical galaxies.

 While we did not investigate barred galaxies, our study of a few
 cases with rapid rotation suggests that in addition to the limited
 range of shapes that elliptical galaxies can assume, stable galaxies
 can only have two ranges of figure rotation $-$ very slow (with
 rotation periods $T_p \gtrsim 5\times10^9$yr) or very fast ($T_p
 \lesssim 2\times 10^8$ yr). 

Since some of the orbits in models with rapid figure rotation can be
quite different from those in models with slow figure rotation, the
implications of our studies could also have consequences for galaxy
and black hole scaling relations. All existing stellar dynamical
measurements of SMBH masses in barred galaxies
(e.g., NGC~1023, NGC~3384, NGC~4151) assume that the dynamics in the
central region is unaffected by the presence of a large-scale bar, and
consequently these systems are modeled as axisymmetric. It has been
argued recently that the ``black hole fundamental plane'' may in fact
be an artifact of the dependence of the $M_{bh}-\sigma$ scaling relation on
host morphology \citep{graham_08b, hu_08}, and in particular could be
a consequence of the fact that barred galaxies appear to be offset
from the $M_{bh}-\sigma$ relation for non-barred galaxies. In addition to
measurement errors in the velocity dispersion parameter $\sigma$ that
arise from large scale streaming motions in the bar
\citep{graham_08a}, neglecting figure rotation could also lead to
errors in $M_{bh}$.

 There are very limited observational prospects for measuring the
 pattern speeds of individual triaxial galaxies and dark matter halos.
 Measurements have only been possible where there is evidence of an
 extended disk-like component in which resonance patterns have been
 excited by the figure rotation or when the Tremaine$-$Weinberg method
 is applicable. Increasing the number of observational measurements of
 pattern speeds in individual galaxies is likely to be
 challenging. Our study of the behavior of orbits suggests
 that triaxial elliptical structures may have only two ranges of pattern
 speeds.
 
However, it is possible that pattern speeds of elliptical galaxies may
be correlated in a statistical sense with observable measures of
angular momentum content such as the $\lambda_R$ parameter used to
separate galaxies into slow and fast rotators
\citep{emsellem_etal_07}.  Certainly, a similar correlation has been found between the pattern speeds  and spin parameter $\lambda$ \citep{peebles_69} of cosmological dark matter halos \citep{bailin_steinmetz_04}.  Further studies of self-consistent models are in progress
 \citep{valluri_11} and will shed light on both the ranges of pattern speeds for which self-consistent equilibrium triaxial galaxies exist, as well as the observable properties. 

%formed in
%cosmological simulations are correlated with the cosmological halobut are not correlated
%with halo mass.  Recent modeling of the internal kinematics of
%galaxies in the SAURON sample has shown that even galaxies which
%appear axisymmetric can have small amounts of triaxiality
%\citep{vandenbosch_dezeeuw_10}.

%Our finding that galaxies with moderate rotation are dominated by
%retrograde short-axis tubes suggests that they could also have large
%streaming motions in a sense opposite to the pattern of the figure. A
%systematic study of the observable properties (such as $\lambda_R$,
%$v/\sigma$, intrinsic luminosity, central cusp slope and surface
%brightness distribution) of large cosmologically motivated sample of
%simulated early type galaxies could be used to search for a
%correlation between observable kinematical parameters and pattern
%speeds. It is likely that a correlation between pattern speed and
%internal measures of kinematics may be found similar to the
%correlation found for dark matter halo simulations. While this will
%still not be evidence for figure rotation in early type galaxies in
%the real Universe, it will likely suggest new avenues for
%statistically estimating the pattern speeds of early type galaxies.

\section*{Acknowledgments}

This work formed A.D.'s undergraduate Honors thesis at the University of
Michigan.

M.V. and A.D. were supported by NSF Grant AST-0908346. D.M. is
supported by Grants AST-0807910 (NSF) and NNX07AH15G (NASA). M.V. thanks Victor Debattista for detailed comments on an earlier
version of this paper and for clarifying numerous properties of
orbits in barred galaxies. MV also thanks Jeremy Bailin for
discussions on figure rotation in cosmological dark matter halos.

%\bibliographystyle{mn2e}
%\bibliographystyle{plain}
%\bibliography{allrefs}
\bibliography{Master}

\end{document}